\documentclass[a4paper,fleqn,onecolumn,usenatbib]{mnras}

\usepackage{amssymb,amsmath,mathrsfs}
\usepackage{multirow}
\usepackage{graphicx}
\usepackage{float}
\usepackage{color}
\usepackage{ulem}
\usepackage{wasysym}
\usepackage{esvect}
\usepackage{newtxtext,newtxmath}
\usepackage[T1]{fontenc}
\usepackage{ae,aecompl}

\newcommand{\cmda}{Celestial Mechanics and Dynamical Astronomy}

\newcommand{\asr}{Advances in Space Research}

\def\etal{\textit{et al. }}

\title{Rotation of a synchronous viscoelastic shell}

\author[B. Noyelles]{Beno\^it Noyelles\thanks{E-mail: benoit.noyelles@unamur.be} \\
University of Namur - Department of Mathematics \& NAmur Center for Complex SYStems (NAXYS) \\
Rempart de la Vierge 8 - B-5000 Namur - Belgium}

\date{Accepted 2017 November 28. Received 2017 November 27; in original form 2017 July 13}

\pubyear{2017}

\begin{document}
\label{firstpage}
\pagerange{\pageref{firstpage}--\pageref{lastpage}}
\maketitle

\begin{abstract}
\par Several natural satellites of the giant planets have shown evidence of a global internal ocean, coated by a thin, icy crust. This crust is probably viscoelastic,
which would alter its rotational response. This response would translate into several rotational quantities, i.e. the obliquity, and the librations at different frequencies,
for which the crustal elasticity reacts differently.
\par This study aims at modelling the global response of the viscoelastic crust. For that, I derive the time-dependency of the tensor of inertia, which I combine with the
time evolution of the rotational quantities, thanks to an iterative algorithm. This algorithm combines numerical simulations of the rotation with a digital filtering 
of the resulting tensor of inertia.
\par The algorithm works very well in the elastic case, provided the problem is not resonant. However, considering tidal dissipation adds different phase lags to the oscillating
contributions, which challenge the convergence of the algorithm.
\end{abstract}


\section{Introduction}

\par Most of the largest satellites of the giant planets are thought to harbour a subsurface ocean under a thin icy shell (Tab.~\ref{tab:shelthickness}).
This is the case for the jovian satellites Europa \citep{pbbcccdfgggghhhkkkmmmmnppsssttww1999,kkrvw2000}, Ganymede \citep{ssbn2002,ss2003,sdrjsfcrmmwnsh2015}
and Callisto \citep{kkssrwp1998,zkk2000}, and for the Saturnian satellite Titan \citep{tgl2005,fgtv2007,bsh2010}. This has also been 
assessed for the mid-sized bodies Dione \citep{brt2016} from its shape and gravity field, Enceladus \citep{tttbjlhp2016} from its longitudinal librations, and possibly for Mimas as well \citep{trlcrrn2014}.
The icy shell that coats this global ocean is thought to have some elasticity; it results from its response to the tidal and rotational deformations
a time-varying shape, that alters the rotational quantities.

\begin{table}
\centering
\caption[Maximal shell thicknesses of satellites of Jupiter and Saturn]{Maximal shell thicknesses of satellites of Jupiter and Saturn, compared with their mean radii. The global ocean for Mimas shoud just be seen as a possible explanation for the observed libration. The mean radius of Ganymede is derived from the solution \emph{Ellipsoid II} of \citet{znohs2015}, which presents the smallest residuals. The limit given for the shell thickness 
of Titan is derived from the topography, while the detection of a Schumann-like resonance in the atmosphere suggests a much thinner crust \citep[e.g.]{bsh2010}.\label{tab:shelthickness}}
\begin{tabular}{l|rr|rr}
Satellite & Shell thickness $\eth$ & Reference & Mean radius $R$ & Reference \\
\hline
J-2 Europa    &  $<35$ km & \citet{ntpm2007}    & $1,560.8$  km & \citet{ntpm2007} \\
J-3 Ganymede  & $<144$ km & \citet{vbcs2014}    & $2,632.97$ km & \citet{znohs2015}\\
S-1 Mimas     &  $<31$ km & \citet{trlcrrn2014} &   $197.49$ km & \citet{trlcrrn2014} \\
S-2 Enceladus &  $<26$ km & \citet{vbt2016}     &   $252.24$ km & \citet{tttbjlhp2016} \\
S-4 Dione     & $<122$ km & \citet{brt2016}     &   $561.4$  km & \citet{t2010} \\
S-6 Titan     & $<200$ km & \citet{hnzi2013}    & $2,574.73$ km & \citet{zshlkl2009} \\
\hline
\end{tabular}
\end{table}

\par \citet{gm2010} alerted the scientific community of the influence of this elasticity on the amplitude of the longitudinal librations.
It generates a restoring torque that tends to counterbalance the effect of the gravitational torque of the parent body, lowering the longitudinal
response, even in considering the couplings with the interior. This prompted \citet{vbt2013} and \citet{rrc2014} to propose a theory of the librations of the Galilean satellites and of Titan considering the
tidal second-order Love number of the shell $k_2$. These studies have shown a reduction of the amplitude of the longitudinal physical librations of 
nearly the same order as the rigid amplitude itself. Considering the elasticity is thus mandatory. In all of these studies, at the most 2 values are considered for the Love numbers,
which correspond to the tides at the zero and diurnal frequencies.

\par I here propose a full numerical model of the rotation of a viscoelastic shell, under the influence of its
own rotation and the gravity field of a perturber. An application will be given for an icy shell that could be the one of Europa, i.e. a
rotating Europa, which response would only be the one of the shell, in neglect of the couplings with the ocean and the core. The full orbital dynamics is considered, and 
a frequency-dependent tidal response is explored. The absence of couplings with the interior reduces the value of the obtained results, in regard to their direct
comparison with future measurements. The goal of the 
paper is a comprehensive treatment of the rotation of a synchronous, viscoelastic body, before the results could be used in further studies considering a realistic interior, i.e. a
viscoelastic icy shell, an ocean, a rigid inner layer, and the mutual interactions associated.

\par In a first part I derive 2 equivalent sets of equations of the problem, in the Hamiltonian and the non-Hamiltonian cases. Then I plug 
a frequency-dependent tidal model, before expressing the observable quantities, i.e. the longitudinal librations and the obliquity, in considering the 
time variations of the inertia of the shell. Finally I present a complete and closed treatment of the elastic problem through a numerical algorithm
which incorporates the three degrees of freedom parameterising the orientation of the shell, and takes into account the perturbations caused by the orbital motion.

\section{Equations of the problem}

\par I model the shell as a body whose tensor of inertia is 

\begin{equation}
\label{eq:inertiatensor}
\mathcal{I}(t)=\left(\begin{array}{ccc}
I_{11}(t) & I_{12}(t) & I_{13}(t) \\
I_{12}(t) & I_{22}(t) & I_{23}(t) \\
I_{13}(t) & I_{23}(t) & I_{33}(t)
\end{array}\right),
\end{equation}
with

\begin{equation}
\label{eq:Iij}
I_{ij} = \iiint\rho\left(r^2\delta_{ij}-r_ir_j\right)\,\textrm{d}V
\end{equation}
the integration being performed over the shell, $\textrm{d}V=\textrm{d}r_1\textrm{d}r_2\textrm{d}r_3$ being an infinitesimal volume element, $\delta_{ij}$ the Kronecker symbol,
and $\rho$ the density. This tensor (\ref{eq:inertiatensor}) is expressed in the figure reference frame $(\hat{f}_1,\hat{f}_2,\hat{f}_3)$ corresponding to the coordinates 
$(r_1,r_2,r_3)$, chosen such that $\hat{f}_1$ is on average collinear with the longest axis, and $\hat{f}_3$ is on average collinear with the shortest one, which is very close to 
the rotation axis. If the shell were rigid, the tensor (\ref{eq:inertiatensor}) would have been constant and the figure frame would have been chosen so that $\mathcal{I}$ be 
diagonal. A viscoelastic shell has a time-varying inertia, that could be written as

\begin{equation}
\label{eq:inertiatv}
\mathcal{I}(t) = <\mathcal{I}>+\delta\mathcal{I}(t)
\end{equation}
with

\begin{equation}
\label{eq:inertiaconstant}
<\mathcal{I}>=\left(\begin{array}{ccc}
<I_{11}> &       0  & 0 \\
0        & <I_{22}> & 0 \\
0        & 0        & <I_{33}>
\end{array}\right)
\end{equation}
and

\begin{equation}
\label{eq:deltainertia}
\delta\mathcal{I}(t)=\left(\begin{array}{ccc}
\delta I_{11}(t) & I_{12}(t)        & I_{13}(t) \\
I_{12}(t)        & \delta I_{22}(t) & I_{23}(t) \\
I_{13}(t)        & I_{23}(t)        & \delta I_{33}(t)
\end{array}\right).
\end{equation}
\par It is a very good approximation to assume that the figure frame $(\hat{f}_1,\hat{f}_2,\hat{f}_3)$ could be obtained by diagonalizing the mean part
of the tensor of inertia, that results in $<\mathcal{I}>$ (Eq.\ref{eq:inertiaconstant}). The time-varying part $\delta\mathcal{I}(t)$ is of a much smaller
order of magnitude, this will be quantified later (Eq.\ref{eq:Imagnitude}). The reader can refer to \citep{m1990} for a more rigourous determination of an optimal reference frame 
for viscoelastic bodies.

\par In the following, I call $(\mathcal{J})=(J_{ij})=(I_{ij})^{-1}$ the invert of the inertia tensor.

\subsection{Geometry of the problem}

\par I consider 3 references frames. The first one, $(\hat{e}_1,\hat{e}_2,\hat{e}_3)$, is centered on the barycenter of the shell and is in translation
with the inertial reference frame used to describe the orbital motion of the body around its parent planet. The second frame $(\hat{n}_1,\hat{n}_2,\hat{n}_3)$
is linked to the angular momentum, and the third one, $(\hat{f}_1,\hat{f}_2,\hat{f}_3)$ is the figure frame as defined above.

\par These reference frames are linked by two sets of Euler's angles. The first one, $(h,K,g)$ locates the position of the angular momentum in the 
inertial frame $(\hat{e}_1,\hat{e}_2,\hat{e}_3)$, while the second one $(g,J,l)$ locates the body frame $(\hat{f}_1,\hat{f}_2,\hat{f}_3)$ with respect to the 
frame $(\hat{n}_1,\hat{n}_2,\hat{n}_3)$. $h$, $K$, $g$, $J$ and $l$ are Andoyer's variables. The angle $K$ is the angle between the angular momentum
and the inertial axis $\hat{e}_3$, it is sometimes called obliquity in the literature, but I will reserve this word for the angle between the angular momentum
and the normal to the instantaneous orbital plane $\varepsilon$. $J$ is the angle between the angular momentum and the figure axis $\hat{f}_3$, it is the amplitude of the
polar motion. We will see later (Fig.~\ref{fig:wob}) that $J$ is in practice negligible, which makes $\varepsilon$ very close to another common definition of the obliquity,
the one associated with the figure axis of our body.

\begin{figure}
\centering
\includegraphics[width=.5\textwidth]{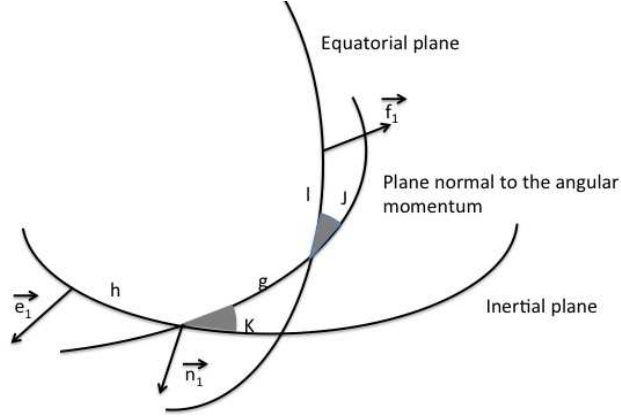}
\caption[The Andoyer variables]{The Andoyer variables and the associated vectors, used in the calculations.}
\end{figure}

\subsection{Hamiltonian formulation}

\par A Hamiltonian formulation of the problem consists in expressing its total energy, i.e. the sum of the kinetic and the potential energies. Here I restrain to the
energy of rotation, the variations of the orbital energy of the body and of its tensor of inertia will appear as a time-dependent distance from its parent planet. 

\subsubsection{The kinetic energy}

\par The kinetic energy $\mathcal{T}$ reads

\begin{equation}
\label{eq:kinetic}
\mathcal{T} = \frac{1}{2} \vec{\omega}\cdot\vec{G},
\end{equation}
where $\vec{\omega}$ is the instantaneous rotation vector, and $\vec{G}$ the angular momentum. These two quantities are linked by the relation

\begin{equation}
\label{eq:defG}
\vec{G} = (\mathcal{I})\cdot\vec{\omega},
\end{equation}
that is equivalent to

\begin{equation}
\label{eq:defomega}
\vec{\omega} = (\mathcal{J})\cdot\vec{G},
\end{equation}
and allows us to write

\begin{equation}
\label{eq:kinetic2}
\mathcal{T} = \frac{1}{2} \vv{(\mathcal{J})\vec{G}}\cdot\vec{G}.
\end{equation}

\par From the definitions of the Euler angles $J$ and $l$, we have

\begin{equation}
\label{eq:angularmomentum}
\vec{G} = G\left(\sin J\sin l\hat{f}_1+\sin J\cos l\hat{f}_2+\cos J\hat{f}_3\right),
\end{equation}
$G$ being the norm of the angular momentum $\vec{G}$, and this yields

\begin{equation}
\label{eq:kineticexpand}
\mathcal{T} = \frac{G^2}{2}\left(J_{11}\sin^2J\sin^2l+J_{22}\sin^2J\cos^2l+J_{33}\cos^2J+J_{12}\sin^2J\sin2l+J_{13}\sin2J\sin l+J_{23}\sin2J\cos l\right).
\end{equation}

\subsubsection{The potential energy\label{sec:potential}}

\par I here express the disturbing potential of a point mass planet $M_p$ on the shell. The planet is located at $P$ whose spherical 
coordinates are $(r,\lambda,\phi)$, $r$ being the distance, $\lambda$ the longitude and $\phi$ the latitude. The mass element 
$\textrm{d}m=\rho\,\textrm{d}r_1\,\textrm{d}r_2\,\textrm{d}r_3$ is located at the point Q$(\tau,l,\theta)$ of the shell, and we have

\begin{eqnarray}
d\mathcal{U}(Q) & = & -\frac{\mathcal{G}M_p\,\textrm{d}m}{QP} \nonumber \\
      & = & -\frac{\mathcal{G}M_p\,\textrm{d}m}{\sqrt{r^2-2r\tau\cos\psi+\tau^2}} \nonumber \\
      & = & -\frac{\mathcal{G}M_p\,\textrm{d}m}{r}\sum_{n=0}^{\infty}\left(\frac{\tau}{r}\right)^nP_n(\cos\psi), \label{eq:dpotential}
\end{eqnarray}
$\psi$ being the angle $\hat{QOP}$, O the barycentre of the shell, and $P_n(x)$ the classical Legendre polynomials. From

\begin{equation}
\label{eq:sphericaltriangle}
\cos\psi = \sin\theta\sin\phi+\cos\theta\cos\phi\cos(\lambda-l)
\end{equation}
we get

\begin{equation}
\label{eq:dpotential2}
d\mathcal{U}(Q) = -\frac{\mathcal{G}M_pdm}{r}\sum_{n=0}^{\infty}\sum_{p=0}^n\alpha_{np}\left(\frac{\tau}{r}\right)^nP_n^p(\sin\phi)P_n^p(\sin\theta)\cos(p(\lambda-l)),
\end{equation}
$P_n^p$ being the associated Legendre functions defined as

\begin{equation}
\label{eq:legendref}
P_n^p(u) = (-1)^p\left(1-u^2\right)^{p/2}\frac{dP_n(u)}{du^p},
\end{equation}

\begin{eqnarray}
\alpha_{n0} & = & 1  \nonumber \\
\alpha_{np} & = & 2\frac{(n-p)!}{(n+p)!}\,\,\, \textrm{if }\,p\ne 0, \label{eq:thealfa}
\end{eqnarray}
and $n\geq p$. Thus, the potential energy $\mathcal{U}$ reads

\begin{equation}
\label{eq:potential}
\mathcal{U} = -\frac{\mathcal{G}M_p}{r}\sum_{n=0}^{\infty}\sum_{p=0}^n\frac{\alpha_{np}}{r^n}P_n^p\left(\sin\phi\right)\left(a_{np}\cos p\lambda+b_{np}\sin p\lambda\right),
\end{equation}
with

\begin{eqnarray}
a_{np} & = & \iiint \tau^n P_n^p(\sin\theta)\cos(pl)\,\textrm{d}m, \label{eq:anp} \\
b_{np} & = & \iiint \tau^n P_n^p(\sin\theta)\sin(pl)\,\textrm{d}m. \label{eq:bnp}
\end{eqnarray}
It stems from this last formula that $b_{n0}=0$.

\par The potential (\ref{eq:potential}) can be rewritten as

\begin{equation}
\label{eq:potentialstokes}
\mathcal{U} = -\frac{\mathcal{G}M_pm}{r}\sum_{n=0}^{\infty}\sum_{p=0}^n\left(\frac{R}{r}\right)^nP_n^p\left(\sin\phi\right)\left(C_{np}\cos(p\lambda)+S_{np}\sin(p\lambda)\right),
\end{equation}
$C_{np}$ and $S_{np}$ being the classical Stokes coefficients, and $R$ the mean outer radius of the shell, i.e. of the satellite. We have straightforwardly

\begin{eqnarray}
C_{np} & = & \alpha_{np}\frac{a_{np}}{mR^n}, \label{eq:Cnp} \\
S_{np} & = & \alpha_{np}\frac{b_{np}}{mR^n},
\end{eqnarray}
and so $S_{n0}=0$.

\par My definition of the internal structure of the shell limits the expansion to the order $n=2$. $n=0$ is the mass $m$, I drop it since it does not
affect the rotational dynamics. The quantities corresponding to $n=1$ are null because the figure reference frame $(\hat{f}_1,\hat{f}_2,\hat{f}_3)$
is centered on the barycentre of the shell, and so we get

\begin{equation}
\label{eq:potentialord2}
\mathcal{U}  = -\frac{\mathcal{G}M_pm}{r}\left(\frac{R}{r}\right)^2\left(C_{20}P_2\left(\sin\phi\right)+P_2^1\left(\sin\phi\right)\left(C_{21}\cos\lambda+S_{21}\sin\lambda\right)+P_2^2\left(\sin\phi\right)\left(C_{22}\cos2\lambda+S_{22}\sin2\lambda\right)\right).
\end{equation}

\par I now introduce the coordinates $(x,y,z)$ of the unit vector pointing to the parent planet from the barycenter of the shell O, i.e. 
$x\hat{f}_1+y\hat{f}_2+z\hat{f}_3=\vv{OP}/OP$. In terms of spherical coordinates this yields

\begin{eqnarray}
x & = & \cos\phi\cos\lambda, \label{eq:xbodyframe} \\
y & = & \cos\phi\sin\lambda, \label{eq:ybodyframe} \\
z & = & \sin\phi. \label{eq:zbodyframe}
\end{eqnarray}

\par From the expressions of the Legendre polynomials and their associated functions

\begin{eqnarray}
P_2(u)   & = & \frac{3}{2}u^2-\frac{1}{2}, \label{eq:P20} \\
P_2^1(u) & = & -3u\sqrt{1-u^2}, \label{eq:P21} \\
P_2^2(u) & = & 3\left(1-u^2\right), \label{eq:P22}
\end{eqnarray}
we have

\begin{eqnarray}
P_2(\sin\phi)               & = & -\frac{3}{2}\left(x^2+y^2\right), \label{eq:pcoef1} \\
P_2^1(\sin\phi)\cos\lambda  & = & -3xz, \label{eq:pcoef2} \\
P_2^1(\sin\phi)\sin\lambda  & = & -3yz, \label{eq:pcoef3} \\
P_2^2(\sin\phi)\cos2\lambda & = & 3\left(x^2-y^2\right), \label{eq:pcoef4} \\
P_2^2(\sin\phi)\sin2\lambda & = & 6xy, \label{eq:pcoef5}
\end{eqnarray}
a constant term having been dropped in Eq.(\ref{eq:pcoef1}), once more because it does not affect the rotational dynamics. The function $P_2^1$ is sometimes given with a positive sign,
because of the omission of the Condon-Shortley phase $(-1)^p$ in the definition of the Lagrange associated function \citep[e.g.]{w2015}. This would result in a different sign for $C_{21}$ and $S_{21}$, as in \citep[e.g.]{cr2013}.

\par The Stokes coefficients become

\begin{eqnarray}
C_{20} & = & \frac{I_{11}+I_{22}-2I_{33}}{2mR^2}, \label{eq:C20} \\
C_{21} & = & \frac{I_{13}}{mR^2}, \label{eq:C21} \\
S_{21} & = & \frac{I_{23}}{mR^2}, \label{eq:S21} \\
C_{22} & = & \frac{I_{22}-I_{11}}{4mR^2}, \label{eq:C22} \\
S_{22} & = & -\frac{I_{12}}{2mR^2}, \label{eq:S22}
\end{eqnarray}
and we can write the disturbing potential as 

\begin{equation}
\label{eq:potentialord2b}
\mathcal{U} = -\frac{3}{2}\mathcal{C}\frac{\mathcal{G}M_p}{r^3}\left(\gamma_1\left(x^2+y^2\right)+\gamma_2xz+\gamma_3yz+\gamma_4\left(x^2-y^2\right)+\gamma_5xy\right),
\end{equation}
with

\begin{eqnarray}
\gamma_1 & = & -C_{20}\frac{mR^2}{\mathcal{C}}=J_2\frac{mR^2}{\mathcal{C}}=\frac{2I_{33}-I_{11}-I_{22}}{2\mathcal{C}}, \label{eq:gamma1} \\
\gamma_2 & = & -2C_{21}\frac{mR^2}{\mathcal{C}}=-2\frac{I_{13}}{\mathcal{C}}, \label{eq:gamma2} \\
\gamma_3 & = & -2S_{21}\frac{mR^2}{\mathcal{C}}=-2\frac{I_{23}}{\mathcal{C}}, \label{eq:gamma3} \\
\gamma_4 & = & 2C_{22}\frac{mR^2}{\mathcal{C}}=\frac{I_{22}-I_{11}}{2\mathcal{C}}, \label{eq:gamma4} \\
\gamma_5 & = & 4S_{22}\frac{mR^2}{\mathcal{C}}=-2\frac{I_{12}}{\mathcal{C}}, \label{eq:gamma5} 
\end{eqnarray}
$\mathcal{C}$ being a normalization factor that will be explicited in Sect.~\ref{sec:timevarying} (Eq.~\ref{eq:Cnorm}). I choose it as very close to the mean value of the polar moment of 
inertia $I_{33}(t)$. In Eq.(\ref{eq:potentialord2b}) the location of the perturber, i.e. the parent planet, appears in the coordinates of the unit vector pointing to
it in the figure frame. I will in fact get its location from ephemerides, that are given in an inertial frame in translation with $(\hat{e}_1,\hat{e}_2,\hat{e}_3)$.
For that I perform 5 rotations that give me the body frame unit coordinates $(x,y,z)$ from the inertial ones $(x_i,y_i,z_i)$, i.e.

\begin{equation}
\label{eq:5rotations}
\left(\begin{array}{c}
x \\
y \\
z \end{array}\right)
= R_3(-l)R_1(-J)R_3(-g)R_1(-K)R_3(-h)\left(\begin{array}{c}
x_i \\
y_i \\
z_i \end{array}\right),
\end{equation}
with 

\begin{equation}
\label{eq:rotation3}
R_3(\phi) = \left(\begin{array}{ccc}
\cos\phi & -\sin\phi & 0 \\
\sin\phi &  \cos\phi & 0 \\
0        &         0 & 1 \end{array}\right)
\end{equation}
and

\begin{equation}
\label{eq:rotation1}
R_1(\phi) = \left(\begin{array}{ccc}
1 &        0 & 0 \\
0 & \cos\phi & -\sin\phi \\
0 & \sin\phi & \cos\phi \end{array}\right).
\end{equation}

The total energy of rotation is then 

\begin{equation}
\label{eq:energyrot}
\mathcal{H}=\mathcal{T}+\mathcal{U}.
\end{equation}

\par I need now to define a set of canonical variables to get a Hamiltonian formulation of the problem. Actually, Andoyer's variables are
canonical:

\begin{center}
$\begin{array}{lll}
l & \hspace{3cm} & L=G\cos J \\
g & \hspace{3cm} & G \\
h & \hspace{3cm} & H=G\cos K,
\end{array}$
\end{center}
$L$ and $H$ being the projections of the angular momentum on $\hat{f}_3$ and $\hat{e}_3$, respectively. Unfortunately, these variables
present two singularities: when $J=0$, i.e. when the angular momentum is colinear to $\hat{f}_3$, $l$ is undefined, while $h$ is undefined
when the angular momentum is colinear to $\hat{e}_3$, i.e. $K=0$. This is the reason why I use the modified Andoyer variables instead:

\begin{center}
$\begin{array}{lll}
p=l+g+h & \hspace{2cm} & P=\frac{G}{n\mathcal{C}} \\
r=-h & \hspace{2cm} & R=\frac{G-H}{n\mathcal{C}}=P(1-\cos K)=2P\sin^2\frac{K}{2} \\
\xi=-\sqrt{2P(1-\cos J)}\sin l & \hspace{2cm} & \eta=\sqrt{2P(1-\cos J)}\cos l, \\
\end{array}$ \\
\end{center}
$n$ being the mean orbital frequency of the satellite. In these variables, the kinetic energy reads

\begin{equation}
\label{eq:kineticanon5}
\mathcal{T} = n\mathcal{C}\frac{P^2}{2}\left(\left(\frac{1}{P}-\frac{\xi^2+\eta^2}{4P^2}\right)\left(J_{11}\xi^2-2J_{12}\xi\eta+J_{22}\eta^2\right)+2\sqrt{\frac{1}{P}-\frac{\xi^2+\eta^2}{4P^2}}\left(1-\frac{\xi^2+\eta^2}{2P}\right)\left(J_{23}\eta-J_{13}\xi\right)+J_{33}\left(1-\frac{\xi^2+\eta^2}{2P}\right)^2\right),
\end{equation}
and the 5 rotations present in the potential energy $\mathcal{U}$ should be expressed with the modified Andoyer's elements as well, and normalized by $n\mathcal{C}$,
to read:

\begin{equation}
\label{eq:potencanon}
\mathcal{U}=-\frac{3}{2}\frac{\mathcal{G}M_p}{nr^3}\left(\gamma_1\left(x^2+y^2\right)+\gamma_2 xz+\gamma_3 yz+\gamma_4 \left(x^2-y^2\right)+\gamma_5 xy\right).
\end{equation}

Then the equations of the problem are the Hamilton equations:

\begin{eqnarray}
\frac{dp}{dt} =  \frac{\partial \mathcal{H}}{\partial P}, & &  \frac{dP}{dt} = -\frac{\partial \mathcal{H}}{\partial p},  \nonumber \\
\frac{dr}{dt} = \frac{\partial \mathcal{H}}{\partial R}, & & \frac{dR}{dt}=-\frac{\partial \mathcal{H}}{\partial r}, \nonumber \\
\frac{d\xi}{dt} = \frac{\partial \mathcal{H}}{\partial \eta}, & & \frac{d\eta}{dt}=-\frac{\partial \mathcal{H}}{\partial \xi}. \label{eq:eqhamilton}
\end{eqnarray}

\par These equations contain time-dependent quantities that are the location of the perturber and the elements of the tensor of inertia $(\mathcal{I})$ and of
its invert $(\mathcal{J})$.

\section{A time-varying inertia tensor}\label{sec:timevarying}

\par The tensor of inertia $(\mathcal{I})$ is controlled by two effects: the rotation of the body, that induces a polar flattening, and the tides raised by the perturber,
that cause the triaxiality. I will first give general formulae for the two effects, before highlighting the role of the rheological model.

\subsection{Tidal and rotational deformations}

\par I model the deformations of the crust following \citet{r2014,rrc2014}, themselves having been inspired by \citet{hla1982}.
I set the radius of a given mass element of the crust as

\begin{equation}
\label{eq:radiushinderer}
r(\tau,\theta,l) = \tau+u_r(\tau,\theta,l),
\end{equation}
with

\begin{equation}
\label{eq:ur}
u_r(\tau,\theta,l) = \tau\left(\sum_{p=0}^2P_2^p\left(\sin\theta\right)\left(d_{2,p}\cos pl+e_{2,p}\sin pl\right)\right).
\end{equation}
The radial deformation $u_r$ comes from

\begin{equation}
\label{eq:urpot}
u_r(\tau,\theta,l) = -\frac{H(\tau)}{g}\delta\mathcal{V},
\end{equation}
$\delta\mathcal{V}$ being the potential resulting from the two deforming effects, $g$ the surface gravity of the body, and $H(\tau)$ a radial function such that $H(R)=h_2$, $h_2$ being the
second-order Love number that characterizes the radial displacement. In the following, I assume that the crust is so thin that $H(\tau)=h_2$. The Eq.(\ref{eq:radiushinderer}) assumes
that the tides act only radially. Actually they also have a lateral action, characterized by the Shida number $l_2$ \citep[e.g.]{mm1975}, and that results in small lateral displacements
$u_{\theta}$ and $u_l$. A straightforward derivation would show that they have only a second-order influence in the expression (\ref{eq:radiushinderer}), this is why I neglect them.
I make the calculation as if $h_2$ were real, in fact the viscoelasticity is partly contained in an imaginary part, this omission will be fixed at the end of this section.

\par The disturbing potential from the parent planet $\delta\mathcal{U}$ is straightforward from the formula (\ref{eq:potential}), i.e.

\begin{equation}
\label{eq:deltapotential}
\delta\mathcal{U} = -\frac{\mathcal{G}M_p}{r}\left(\frac{R}{r}\right)^2\sum_{p=0}^2\alpha_{2p}P_2^p(\sin\theta)P_2^p(\sin\phi)\left(\cos p\lambda\cos pl+\sin p\lambda\sin pl\right),
\end{equation}
with $\alpha_{20}=1$, $\alpha_{21}=1/3$ and $\alpha_{22}=1/12$ (Eq.\ref{eq:thealfa}).

\par The potential due to the rotation reads \citep{jv2014}

\begin{equation}
\label{eq:potentialrot}
\delta\mathcal{R} = -R^2\left(\frac{\omega_3^2}{2}\left(r_1^2+r_2^2\right)+\frac{\omega_2^2}{2}\left(r_1^2+r_3^2\right)+\frac{\omega_1^2}{2}\left(r_2^2+r_3^2\right)-\omega_1\omega_3r_1r_3-\omega_1\omega_2r_1r_2-\omega_2\omega_3r_2r_3\right),
\end{equation}
this can be straightforwardly obtained from the expression of the centrifugal force $\vec{\omega}\times\left(\vec{\omega}\times\vec{r}\right)$. 
As shown, e.g., in \citet[App~B.3]{e2012}, the centrifugal force and the ensuing deformation comprise a purely radial component and a component mimicking
a quadrupole tidal deformation. In using the relation $r_1^2+r_2^2+r_3^2=1$ and the formulae (\ref{eq:pcoef1}) to (\ref{eq:pcoef5}), $\delta\mathcal{R}$ can be written as

\begin{equation}
\label{eq:potentialrot25}
\delta\mathcal{R} = -R^2\left(-\frac{1}{3}P_2(\sin\theta)\left(\omega_3^2-\frac{\omega_1^2+\omega_2^2}{2}\right)+\frac{\omega_2^2-\omega_1^2}{12}P_2^2\cos 2l+P_2^1(\sin\theta)\left(\frac{\omega_1\omega_3}{3}\cos l+\frac{\omega_2\omega_3}{3}\sin l\right)-\frac{\omega_1\omega_2}{6}P_2^2(\sin\theta)\sin 2l\right).
\end{equation}

\par From $\delta\mathcal{V}=\delta\mathcal{U}+\delta\mathcal{R}$ and the Eq.(\ref{eq:ur}) and (\ref{eq:urpot}), I get

\begin{eqnarray}
d_{20} & = & -\frac{h_2}{g\tau}\left(\frac{R^2}{3}\left(\omega_3^2-\frac{\omega_1^2+\omega_2^2}{2}\right)-\frac{\mathcal{G}M_p}{r}\left(\frac{R}{r}\right)^2\left(z^2-\frac{x^2+y^2}{2}\right)\right), \label{eq:d20b} \\
d_{21} & = & \frac{h_2}{g\tau}\left(\frac{R^2}{3}\omega_1\omega_3-\frac{\mathcal{G}M_p}{r}\left(\frac{R}{r}\right)^2xz\right), \label{eq:d21b} \\
e_{21} & = & \frac{h_2}{g\tau}\left(\frac{R^2}{3}\omega_2\omega_3-\frac{\mathcal{G}M_p}{r}\left(\frac{R}{r}\right)^2yz\right), \label{eq:e21b} \\
d_{22} & = & -\frac{h_2}{g\tau}\left(\frac{R^2}{12}\left(\omega_1^2-\omega_2^2\right)-\frac{\mathcal{G}M_p}{4r}\left(\frac{R}{r}\right)^2\left(x^2-y^2\right)\right), \label{eq:d22b} \\
e_{22} & = & -\frac{h_2}{g\tau}\left(\frac{R^2}{6}\omega_1\omega_2-\frac{\mathcal{G}M_p}{2r}\left(\frac{R}{r}\right)^2xy\right). \label{eq:e22b}
\end{eqnarray}

The definitions of the moments of inertia (Eq.\ref{eq:Iij}) give

\begin{equation}
\label{eq:Iabint}
I_{\alpha\beta} = \iiint \rho r^4 \mathcal{F}_{\alpha\beta}(\theta,l)\,\textrm{d}r\,\textrm{d}\theta\,\textrm{d}l
\end{equation}

with

\begin{eqnarray}
\mathcal{F}_{11}(\theta,l) & = & (\cos^2\theta\sin^2 l+\sin^2\theta)\cos\theta, \label{eq:F11} \\
\mathcal{F}_{12}(\theta,l) & = & -\frac{1}{2}\cos^3\theta\sin 2l, \label{eq:F12} \\
\mathcal{F}_{13}(\theta,l) & = & -\cos^2\theta\sin\theta\cos l, \label{eq:F13} \\
\mathcal{F}_{22}(\theta,l) & = & (\cos^2\theta\cos^2 l+\sin^2\theta)\cos\theta, \label{eq:F22} \\
\mathcal{F}_{23}(\theta,l) & = & -\cos^2\theta\sin\theta\sin l, \label{eq:F23} \\
\mathcal{F}_{33}(\theta,l) & = & \cos^3\theta. \label{eq:F33}
\end{eqnarray}

\par I now make a change of variables to use $\tau$ instead of $r$, following the Eq.(\ref{eq:radiushinderer}). This is a switch from the Lagrange description, i.e. with respect to the initial
coordinates of the considered mass element, to the Euler one, where the mass element is identified with its final coordinates. It yields

\begin{equation}
\label{eq:dradiushinderer}
\frac{\textrm{d}r}{\textrm{d}\tau} = \frac{r}{\tau}+\tau\sum_{j=0}^2P_2^j(\sin\theta)\left(\frac{\textrm{d}d_{2,j}}{\textrm{d}\tau}\cos jl+\frac{\textrm{d}e_{2,j}}{\textrm{d}\tau}\sin jl\right),
\end{equation}
and from the linearization

\begin{equation}
\label{eq:linear}
\frac{\textrm{d}r}{\textrm{d}\tau} = 5\left(\tau^4+\frac{\textrm{d}u_r\tau^5}{\textrm{d}\tau}\right),
\end{equation}
I get

\begin{eqnarray}
I_{\alpha\beta} & = & \int_{\tau=R-\eth}^{\tau=R}\int_{\theta=-\frac{\pi}{2}}^{\theta=\frac{\pi}{2}}\int_{l=0}^{l=2\pi}\rho\left[\tau^4+\frac{\textrm{d}(d_{20}\tau^5)}{\textrm{d}\tau}P_2(\sin\theta)+\frac{\textrm{d}(d_{21}\tau^5)}{\textrm{d}\tau}P_2^1(\sin\theta)\cos l+\frac{\textrm{d}(e_{21}\tau^5)}{\textrm{d}\tau}P_2^1(\sin\theta)\sin l\right. \nonumber \\
& & \left.+\frac{\textrm{d}(d_{22}\tau^5)}{\textrm{d}\tau}P_2^2(\sin\theta)\cos 2l+\frac{\textrm{d}(e_{22}\tau^5)}{\textrm{d}\tau}P_2^2(\sin\theta)\sin 2l\right]\mathcal{F}_{\alpha\beta}(\theta,l)\,\textrm{d}\tau\,\textrm{d}\theta\,\textrm{d}l, \label{eq:bigint}
\end{eqnarray}
$\eth$ being the thickness of the crust. This gives, in using the Eq.(\ref{eq:d20b}) to (\ref{eq:e22b}):

\begin{eqnarray}
I_{11} & = & \frac{8\pi}{3}\rho X_4-\frac{16\pi}{45}\rho\frac{h_2}{g}\left(\omega_3^2+\omega_2^2-2\omega_1^2\right)R^2X_3+\frac{16\pi}{15}\rho\frac{h_2}{g}\frac{\mathcal{G}M_p}{r}\left(\frac{R}{r}\right)^2X_3\left(z^2-2x^2+y^2\right), \label{eq:I11cart}  \\
I_{12} & = & \frac{16\pi}{15}\rho\frac{h_2}{g}R^2X_3\omega_1\omega_2-\frac{16\pi}{5}\rho\frac{h_2}{g}\frac{\mathcal{G}M_p}{r}\left(\frac{R}{r}\right)^2X_3xy, \label{eq:I12cart} \\
I_{13} & = & \frac{16\pi}{15}\rho\frac{h_2}{g}R^2X_3\omega_1\omega_3-\frac{16\pi}{5}\rho\frac{h_2}{g}\frac{\mathcal{G}M_p}{r}\left(\frac{R}{r}\right)^2X_3xz, \label{eq:I13cart} \\
I_{22} & = & \frac{8\pi}{3}\rho X_4-\frac{16\pi}{45}\rho\frac{h_2}{g}\left(\omega_3^2+\omega_1^2-2\omega_2^2\right)R^2X_3+\frac{16\pi}{15}\rho\frac{h_2}{g}\frac{\mathcal{G}M_p}{r}\left(\frac{R}{r}\right)^2X_3\left(z^2+x^2-2y^2\right), \label{eq:I22cart} \\
I_{23} & = & \frac{16\pi}{15}\rho\frac{h_2}{g}R^2X_3\omega_2\omega_3-\frac{16\pi}{5}\rho\frac{h_2}{g}\frac{\mathcal{G}M_p}{r}\left(\frac{R}{r}\right)^2X_3yz, \label{eq:I23cart} \\
I_{33} & = & \frac{8\pi}{3}\rho X_4-\frac{16\pi}{45}\rho\frac{h_2}{g}\left(\omega_1^2+\omega_2^2-2\omega_3^2\right)R^2X_3+\frac{16\pi}{15}\rho\frac{h_2}{g}\frac{\mathcal{G}M_p}{r}\left(\frac{R}{r}\right)^2X_3\left(x^2+y^2-2z^2\right), \label{eq:I33cart}
\end{eqnarray}
with

\begin{equation}
\label{eq:Xn}
X_n = \int_{R-\eth}^R \tau^n\,\textrm{d}\tau \approx \eth R^n.
\end{equation}

\par These formulae are valid for any constant rotation rate. A nonuniform rotation actually induces toroidal deformation, which is neglected here since it is a high-order effect for a near-spherical body.
It is assumed that the Love number $h_2$ is constant over the thickness of the crust, and that it does not depend on the frequency
of the excitation. In this paper, I assume that the shell is thin enough for $h_2$ to be independent of the depth. However, the assumption that $h_2$ is frequency-independent is physically 
unjustified. In the studies assuming that the body is rigid, $h_2$ is assumed to be null except for secular excitations. 
This results in a frozen shape, that is triaxial for despun bodies like synchronous planetary satellites, or Mercury. On the contrary, considering $h_2$ to be frequency-independent would overestimate
the role of the elasticity in the crustal response to any external perturbation. In this study I assume $h_2=h_2(\chi)$, where $\chi$ is the tidal frequency, defined as the absolute value of any frequency 
of tidal excitation.

\par Relaxing this approximation requires to be more specific on the rotational dynamics. This paper addresses the rotation of tidally despun, synchronous, planetary satellites. In this analytical
study, I neglect the polar motion, i.e. $\omega_1=\omega_2=0$. Moreover I assume that the satellite is perturbed only by its parent planet, and that its orbit is a uniformly precessing ellipse. Its
orbital elements are the semimajor axis $a$, the eccentricity $e$, the inclination $I$, the mean longitude $\lambda$, the argument of the pericentre $\varpi$, and the argument of the ascending node $\ascnode$.
These elements are defined in an inertial reference frame, usually linked to the equator of the planet at J2000. These assumptions imply that $a$, $e$ and $I$ are constant, and that $\lambda$, $\varpi$
and $\ascnode$ have constant time derivatives, written as $\dot{\lambda}=n$, $\dot{\varpi}$ and $\dot{\ascnode}$, respectively.

\par Here are the steps of the expansion of the moments of inertia with respect to the orbital elements:

\begin{enumerate}

\item Substitution of $\omega_3$ by $n$. This means that I neglect, in the framework of this analytical study, the physical librations, due to the triaxiality of the body. I consider only
the optical librations (see Sect.\ref{sec:libra}). This assumption is relaxed in the numerical study (Sect. \ref{sec:numerical}).

\item Substitution of $\mathcal{G}M_p$ by $n^2a^3$, i.e. the mass of the satellite is neglected with respect to the mass of the parent planet in the orbital motion.

\item Expression of the quantities $x$, $y$, $z$ with the rotations

\begin{equation}
\label{eq:exprxyz}
\left(\begin{array}{c}
x \\
y \\
z \end{array}\right) = R_3(-l-g)R_1(-K)R_3(\ascnode-h)R_1(i)R_3(\varpi-\ascnode)\left(\begin{array}{c}
\cos f \\
\sin f \\
0 \end{array}\right),
\end{equation}
where $f$ is the true anomaly.

\item Once these expressions (\ref{eq:exprxyz}) have been inserted into the expression of the moments of inertia, they can be expanded with respect to the eccentricity in using the classical relations

\begin{eqnarray}
\frac{a}{r} & = & 1+2\sum_{\nu=1}^{\infty} J_{\nu}(\nu e)\cos (\nu(\lambda-\varpi)), \label{eq:asr} \\
\cos f & = & 2\frac{1-e^2}{e}\sum_{\nu=1}^{\infty} J_{\nu}(\nu e)\cos (\nu(\lambda-\varpi))-e, \label{eq:cosf} \\
\sin f & = & 2\sqrt{1-e^2}\sum_{\nu=1}^{\infty}\frac{\textrm{d}J_{\nu}(\nu e)}{\textrm{d}e}\frac{\sin(\nu(\lambda-\varpi))}{\nu}, \label{eq:sinf}
\end{eqnarray}
where $J_{\nu}$ are the Bessel functions of the first kind \citep[e.g.]{dl2013}.

\par I now use dynamical properties of the despun, synchronous rotation (see e.g. \citet{nlv2008}). I introduce the resonant argument $\sigma=l+g+h-\lambda+\pi$ that 
averages to $0$ for a synchronous body. There are oscillations around $0$, i.e. forced librations in longitude, but their amplitude is neglected here, since they are proportional to the triaxiality. Moreover, 
it stems from the Third Cassini Law \citep{c1693,c1966,n2009} that $\ascnode=h$ and $\varepsilon=K-I$, $\varepsilon$ being the instantaneous obliquity.

\end{enumerate}

\par After expansion with respect to the small quantities $e$ and $\varepsilon$, we have, at the second order:

\begin{eqnarray}
I_{11} & = & \frac{8\pi}{3}\rho X_4+\frac{n^2R^2}{g}\pi\rho X_3\left(h_2(\chi_0)\left(-\frac{112}{45}+\frac{16}{5}e^2+\frac{8}{5}\varepsilon^2\right)-\frac{32}{5}h_2(\chi_1)e\cos(\lambda-\varpi)\right. \nonumber \\
& & \left.-16h_2(\chi_3)e^2\cos 2(\lambda-\varpi)-\frac{8}{5}h_2(\chi_4)\varepsilon^2\cos2(\lambda-\ascnode)\right), \label{eq:I11freq} \\
I_{12} & = & \frac{4n^2R^2}{5g}\pi\rho X_3\left(-8h_2(\chi_1)e\sin(\lambda-\varpi)-17h_2(\chi_3)e^2\sin2(\lambda-\varpi)+h_2(\chi_4)\varepsilon^2\sin2(\lambda-\ascnode)\right), \label{eq:I12freq} \\
I_{13} & = & -\frac{n^2R^2}{g}\pi\rho X_3\varepsilon\left(\frac{16}{5}h_2(\chi_2)\sin(\lambda-\ascnode)+8h_2(\chi_5)e\sin (2\lambda-\varpi-\ascnode)+\frac{8}{5}h_2(\chi_6)e\sin(\varpi-\ascnode)\right), \label{eq:I13freq} \\
I_{22} & = & \frac{8\pi}{3}\rho X_4+\frac{n^2R^2}{g}\pi\rho X_3\left(h_2(\chi_0)\left(\frac{32}{45}-\frac{24}{5}e^2\right)+\frac{16}{5}h_2(\chi_1)e\cos (\lambda-\varpi)+\frac{56}{5}h_2(\chi_3)e^2\cos 2(\lambda-\varpi)\right), \label{eq:I22freq} \\
I_{23} & = & \frac{16}{5}\frac{n^2R^2}{g}\pi\rho X_3 e \varepsilon\left(h_2(\chi_5)\cos(2\lambda-\varpi-\ascnode)-h_2(\chi_6)\cos(\varpi-\ascnode)\right), \label{eq:I23freq} \\
I_{33} & = & \frac{8\pi}{3}\rho X_4+\frac{n^2R^2}{g}\pi\rho X_3\left(h_2(\chi_0)\left(\frac{16}{9}+\frac{8}{5}e^2-\frac{8}{5}\varepsilon^2\right)+\frac{16}{5}h_2(\chi_1)e\cos(\lambda-\varpi)\right. \nonumber \\
& & \left.+\frac{24}{5}h_2(\chi_3)e^2\cos 2(\lambda-\varpi)+\frac{8}{5}h_2(\chi_4)\varepsilon^2\cos 2(\lambda-\ascnode)\right), \label{eq:I33freq}
\end{eqnarray}
where $\chi_i$ are tidal frequencies. 

\par Up to now I have assumed the $h_2(\chi)$ to be real numbers. In fact they are complex, and their imaginary part contain the quality factor, or dissipation function $Q_h$, as 
$Im(h_2^*)=-h_2/Q_h$. The dissipation function is often defined with the tidal Love number $k_2$ and denoted $Q$, here I used a definition for $h_2$ present in \citep{wb2015}.
As shown in \textit{Ibid.}, this imaginary part is responsible for a phase shift in the response of the crust, and it can be considered in replacing in the formulae (\ref{eq:I11freq})
to (\ref{eq:I33freq}), $h_2\cos$ by $h_2\cos+h_2/Q_h\sin$ and $h_2\sin$ by $h_2\sin-h_2/Q_h\cos$. This comes from the definition of the phase lag $\epsilon_2(\chi)$ as

\begin{equation}
\label{eq:phaselag}
\sin\epsilon_2(\chi) = -\frac{Im(h_2^*(\chi))}{|h_2^*(\chi)|}
\end{equation}
and

\begin{equation}
\label{eq:Qh}
Q_h^{-1}\equiv |\sin\epsilon_2(\chi)|,
\end{equation}
from which we have

\begin{eqnarray}
h_2(\chi)\cos(...-\epsilon_2(\chi)) & = & h_2(\chi)[\cos(...)\cos\epsilon_2(\chi)+\sin(...)\sin\epsilon_2(\chi)] \nonumber \\
& \approx & h_2(\chi)[\cos(...)+\frac{\sin(...)}{Q_h(\chi)}]. \label{eq:newh2cos}
\end{eqnarray}
and

\begin{eqnarray}
h_2(\chi)\sin(...-\epsilon_2(\chi)) & = & h_2(\chi)[\sin(...)\cos\epsilon_2(\chi)-\cos(...)\sin\epsilon_2(\chi)] \nonumber \\
& \approx & h_2(\chi)[\sin(...)+\frac{\cos(...)}{Q_h(\chi)}]. \label{eq:newh2sin}
\end{eqnarray}

This yields:

\begin{eqnarray}
I_{11} & = & \frac{8\pi}{3}\rho X_4+\frac{n^2R^2}{g}\pi\rho X_3\left(h_2(\chi_0)\left(-\frac{112}{45}+\frac{16}{5}e^2+\frac{8}{5}\varepsilon^2\right)-\frac{32}{5}e\left(h_2(\chi_1)\cos(\lambda-\varpi)+\left(\frac{h_2}{Q_h}\right)(\chi_1)\sin(\lambda-\varpi)\right)\right. \nonumber \\
& & \left.-16e^2\left(h_2(\chi_3)\cos 2(\lambda-\varpi)+\left(\frac{h_2}{Q_h}\right)(\chi_3)\sin 2(\lambda-\varpi)\right)\right. \nonumber \\
& & \left.-\frac{8}{5}\varepsilon^2\left(h_2(\chi_4)\cos 2(\lambda-\ascnode)+\left(\frac{h_2}{Q_h}\right)(\chi_4)\sin 2(\lambda-\ascnode)\right)\right), \label{eq:I11qfreq} \\
I_{12} & = & \frac{4n^2R^2}{5g}\pi\rho X_3\left(-8e\left(h_2(\chi_1)\sin(\lambda-\varpi)-\left(\frac{h_2}{Q_h}\right)(\chi_1)\cos(\lambda-\varpi)\right)\right. \nonumber \\
& & \left.-17e^2\left(h_2(\chi_3)\sin2(\lambda-\varpi)-\left(\frac{h_2}{Q_h}\right)(\chi_3)\cos2(\lambda-\varpi)\right)+\varepsilon^2\left(h_2(\chi_4)\sin2(\lambda-\ascnode)-\left(\frac{h_2}{Q_h}\right)(\chi_4)\cos2(\lambda-\ascnode)\right)\right), \label{eq:I12qfreq} \\
I_{13} & = & -\frac{n^2R^2}{g}\pi\rho X_3\varepsilon\left(\frac{16}{5}\left(h_2(\chi_2)\sin(\lambda-\ascnode)-\left(\frac{h_2}{Q_h}\right)(\chi_2)\cos(\lambda-\ascnode)\right)\right. \nonumber \\
& & \left.+8e\left(h_2(\chi_5)\sin (2\lambda-\varpi-\ascnode)-\left(\frac{h_2}{Q_h}\right)(\chi_5)\cos (2\lambda-\varpi-\ascnode)\right)\right. \nonumber \\
& & \left.+\frac{8}{5}e\left(h_2(\chi_6)\sin(\varpi-\ascnode)-\left(\frac{h_2}{Q_h}\right)(\chi_6)\cos(\varpi-\ascnode)\right)\right), \label{eq:I13qfreq} \\
I_{22} & = & \frac{8\pi}{3}\rho X_4+\frac{n^2R^2}{g}\pi\rho X_3\left(h_2(\chi_0)\left(\frac{32}{45}-\frac{24}{5}e^2\right)+\frac{16}{5}e\left(h_2(\chi_1)\cos (\lambda-\varpi)+\left(\frac{h_2}{Q_h}\right)(\chi_1)\sin (\lambda-\varpi)\right)\right. \nonumber \\
& & \left.+\frac{56}{5}e^2\left(h_2(\chi_3)\cos 2(\lambda-\varpi)+\left(\frac{h_2}{Q_h}\right)(\chi_3)\sin 2(\lambda-\varpi)\right)\right), \label{eq:I22qfreq} \\
I_{23} & = & \frac{16}{5}\frac{n^2R^2}{g}\pi\rho X_3 e \varepsilon\left(h_2(\chi_5)\cos(2\lambda-\varpi-\ascnode)+\left(\frac{h_2}{Q_h}\right)(\chi_5)\sin(2\lambda-\varpi-\ascnode)\right. \nonumber \\
& &  \left.-h_2(\chi_6)\cos(\varpi-\ascnode)-\left(\frac{h_2}{Q_h}\right)(\chi_6)\sin(\varpi-\ascnode)\right), \label{eq:I23qfreq} \\
I_{33} & = & \frac{8\pi}{3}\rho X_4+\frac{n^2R^2}{g}\pi\rho X_3\left(h_2(\chi_0)\left(\frac{16}{9}+\frac{8}{5}e^2-\frac{8}{5}\varepsilon^2\right)+\frac{16}{5}e\left(h_2(\chi_1)\cos(\lambda-\varpi)+\left(\frac{h_2}{Q_h}\right)(\chi_1)\sin(\lambda-\varpi)\right)\right. \nonumber \\
& & \left.+\frac{24}{5}e^2\left(h_2(\chi_3)\cos 2(\lambda-\varpi)+\left(\frac{h_2}{Q_h}\right)(\chi_3)\sin 2(\lambda-\varpi)\right)\right. \nonumber \\
& & \left.+\frac{8}{5}\varepsilon^2\left(h_2(\chi_4)\cos 2(\lambda-\ascnode)+\left(\frac{h_2}{Q_h}\right)(\chi_4)\sin 2(\lambda-\ascnode)\right)\right). \label{eq:I33qfreq}
\end{eqnarray}

In these last expressions, $h_2$ is the real part of the complex Love number $h_2^*$, and $h_2/Q_h$ is the opposite of its imaginary part.
These expressions emphasize the role of different frequencies, that are

\begin{eqnarray}
\chi_0 & = & 0, \label{eq:chi0} \\
\chi_1 & = & \left|n-\dot{\varpi}\right|, \label{eq:chi1} \\
\chi_2 & = & \left|n-\dot{\ascnode}\right|, \label{eq:chi2} \\
\chi_3 & = & 2\left|n-\dot{\varpi}\right|, \label{eq:chi3} \\
\chi_4 & = & 2\left|n-\dot{\ascnode}\right|, \label{eq:chi4} \\
\chi_5 & = & \left|2n-\dot{\varpi}-\dot{\ascnode}\right|, \label{eq:chi5} \\
\chi_6 & = & \left|\dot{\varpi}-\dot{\ascnode}\right|. \label{eq:chi6}
\end{eqnarray}
At the secular frequency $\chi_0$, we have $(h_2/Q_h)(\chi_0) = 0$.

\par These formulae first give an order of magnitude of the impact of the viscoelasticity on the tensor of inertia. For the Eqs.(\ref{eq:I11qfreq}) to (\ref{eq:I33qfreq}) we can write

\begin{equation}
\label{eq:Imagnitude}
(\mathcal{I}) \propto \left(\begin{array}{ccc}
<I_{11}>    & e            & \varepsilon \\
e           & <I_{22}>     & e\varepsilon \\
\varepsilon & e\varepsilon & <I_{33}>
\end{array}\right),
\end{equation}
i.e. the viscoelasticity is a small effect, without which the tensor of inertia can be written under a diagonal form. The quantities $I_{12}$ and $I_{13}$ are first-order quantities of the eccentricity 
and the obliquity respectively, we will see later (Sect.\ref{sec:libra}) that they are linked with the longitudinal and latitudinal librations. $I_{23}$ is a second-order quantity.

\par We can also see different orders of magnitudes in the tidal frequencies. $\chi_1$ and $\chi_2$ are of the order of the orbital frequency $n$, $\chi_3$, $\chi_4$ and $\chi_5$ are of the order of $2n$,
while $\chi_6$ is much smaller since it is related to the precession of the orbit, and $\chi_0$ addresses the mean shape of the body. The dependency of the complex Love number $h_2^*$ on the tidal frequency is 
discussed in the next section.

\par From the Eq.~\ref{eq:I33qfreq}, the normalization factor $\mathcal{C}$ introduced in Sect.~\ref{sec:potential} reads

\begin{equation}
\label{eq:Cnorm}
\mathcal{C} = \frac{8\pi}{3}\rho X_4+\frac{n^2R^2}{g}\pi\rho X_3h_2(\chi_0)\left(\frac{16}{9}+\frac{8}{5}e^2-\frac{8}{5}\varepsilon^2\right).
\end{equation}

\subsection{Influence of the rheology}

\par There are several models of rheology in the literature, modeling the way the tidal coefficients depend on the excitation frequency $\chi$. The reader can find a comprehensive
explanation in \citep{mc2013}. Here I will just address the classical Maxwell model, and the Andrade one, which is probably more realistic for planetary ices.

\par The tidal models usually give the frequency behavior of the tidal Love coefficient $k_2$, while I have used $h_2$ above. Another difference is that $k_2$ is usually implicitly
given at its main tidal frequency, which depends on its orbital and spin periods. For synchronous satellites, it is actually the orbital frequency, as suggested by the Eqs.(\ref{eq:I11freq})
to (\ref{eq:I33freq}).

\par At the zero frequency $\chi_0$, $h_2$ and $k_2$ are known as secular Love numbers. $h_2(\chi_0)$ verifies the following formulae, which are derived from the Radau equation \citep[e.g.]{j1952}:

\begin{eqnarray}
c & = & b-\frac{h_2(\chi_0)}{2}qR, \label{eq:chydro} \\
a & = & b+\frac{3}{2}qh_2(\chi_0)R, \label{eq:ahydro}
\end{eqnarray}
with

\begin{eqnarray}
q & = & \frac{n^2R^3}{\mathcal{G}m}, \label{eq:qhydro}
\end{eqnarray}
$n$ being the orbital motion, while $k_2(\chi_0)$ verifies \citep{rbga1997}:

\begin{equation}
\label{eq:C22t}
C_{22} = k_2(\chi_0)\frac{q}{4}.
\end{equation}

\par I compute 33,935 model-Europas in Sect.~\ref{sec:interior}, which give

\begin{eqnarray}
h_2(\chi_0) & = & 2.05513 \pm 0.01152, \nonumber \\
k_2(\chi_0) & = & 1.05554 \pm 0.01154, \nonumber \\
1+k_2(\chi_0)-h_2(\chi_0) & = & 4.109\times10^{-4} \pm 1.278\times10^{-4}, \nonumber
\end{eqnarray}
the given uncertainties being actually standard deviations.

\par We can see that $h_2(\chi_0)-k_2(\chi_0)$ is close to 1, which is a classical relation, at least at the orbital frequency. The departure from this equality does not show any obvious
relation with $\eth/R$, contrary to the suggestion of \citet[Eq.~3]{wzsl2006}. In the following, I will assume the quantity $k_2-h_2$ to be constant with respect to the tidal frequency $\chi$,
which gives

\begin{equation}
\label{eq:thinshellmoi}
h_2(\chi) = k_2(\chi)+h_2(\chi_0)-k_2(\chi_0).
\end{equation}
$k_2(\chi)$ is the modulus of the complex number \citep[e.g.]{e2012}

\begin{equation}
\label{eq:k2general}
k_2^*(\chi) = k_2(\chi_0)\frac{J^*(\chi)}{J^*(\chi)+A_2/\mu},
\end{equation}
where $A_2$ is a coefficient defined by

\begin{equation}
\label{eq:A2}
A_2 = \frac{19}{2}\frac{\mu}{\rho g R},
\end{equation}
and $J^*(\chi)$ is the complex compliance.

\par The Maxwell model is probably the most classical of the rheological models, it represents a viscoelastic behavior in which the elasticity and the viscosity are not coupled.
It depends on only one parameter, the Maxwell time $\tau_M$, itself defined as the ratio between the viscosity $\eta$ and the rigidity $\mu$:

\begin{equation}
\label{eq:Maxwelltime}
\tau_M = \frac{\eta}{\mu},
\end{equation}
and we have \citep[Tab.~3.2]{k2008}

\begin{equation}
\label{eq:Maxwellcompliance}
J^*(\chi) = \frac{\chi\tau_M-\imath}{\mu\chi\tau_M},
\end{equation}
yielding \citep[e.g.]{e2015}

\begin{eqnarray}
\mathfrak{R}(k_2^*) & = & k_2(\chi_0)\frac{1+(1+A_2)\chi^2\tau_M^2}{1+(1+A_2)^2\chi^2\tau_M^2}, \label{eq:Rek2Maxwell} \\
-\mathfrak{I}(k_2^*) & = & k_2(\chi_0)\frac{A_2\chi\tau_M}{1+(1+A_2)^2\chi^2\tau_M^2}. \label{eq:Imk2Maxwell}
\end{eqnarray}

\par At high frequencies, the Maxwell model lacks of accuracy because it does not render an expected anelastic behavior, i.e. a non-instantaneous but reversible deformation, at least 
when dissipation is neglected. The Andrade model is there more appropriate. This is the reason why \citet{e2012} proposed the so-called Andrade-Maxwell model, that corresponds to the Andrade model 
at high frequencies and to the Maxwell model at lower frequencies. Its complex compliance reads:

\begin{equation}
\label{eq:Efroimskycompliance}
\bar{J}(\chi) = \left(1+\left(\imath\chi\tau_A\right)^{-\alpha}\Gamma(1+\alpha)-\imath\left(\chi\tau_M\right)^{-1}\right)/\mu,
\end{equation}
$\Gamma$ being the classical $\Gamma$ function defined as

\begin{equation}
\label{eq:Gamma}
\Gamma(1+\alpha) = \int_0^{+\infty} z^\alpha e^{-z}\,\textrm{d}z.
\end{equation}

\par We can see that this model depends on 3 tidal parameters, that are the Maxwell time $\tau_M$ (Eq.\ref{eq:Maxwelltime}), an Andrade time $\tau_A$ that has been introduced by 
\citet[Eq.~78]{e2012}, and an Andrade parameter $\alpha$. The Andrade time should be equal to the Maxwell time to have a continuous transition between viscoelasticy and 
anelasticity. $\alpha$ is usually assumed to lie between 0.1 and 0.5, I will take $\alpha=0.3$ as it is classically assumed for outer layers. For the solid Earth as a
whole, this parameter takes a lower value of $\alpha\approx0.18$, owing to the presence of partial melts in the mantle. The resulting expressions for the Andrade-Maxwell model are \citep[e.g.]{m2012}

\begin{eqnarray}
\mathfrak{R}(k_2^*) & = & k_2(\chi_0)\frac{\mathcal{A}^2+\mathcal{A}A_2+\mathcal{B}^2}{\mathcal{A}^2+2\mathcal{A}A_2+A_2^2+\mathcal{B}^2}, \label{eq:Rek2Efroimsky} \\
-\mathfrak{I}(k_2^*) & = & k_2(\chi_0)\frac{A_2\mathcal{B}}{\mathcal{A}^2+2\mathcal{A}A_2+A_2^2+\mathcal{B}^2}, \label{eq:Imk2Efroimsky}
\end{eqnarray}
with

\begin{eqnarray}
\mathcal{A} & = & 1+\left(\chi\tau_A\right)^{-\alpha}\cos\left(\frac{\alpha\pi}{2}\right)\Gamma(1+\alpha), \label{eq:Aronde} \\
\mathcal{B} & = & \left(\chi\tau_M\right)^{-1}+\left(\chi\tau_A\right)^{-\alpha}\sin\left(\frac{\alpha\pi}{2}\right)\Gamma(1+\alpha). \label{eq:Bronde}
\end{eqnarray}

\par The Andrade parameter $\tau_A$ is an order of magnitude of the period above which the excitation will generate the Andrade creep, responsible for anelasticity. Setting $\tau_A$ to the 
infinity renders the Maxwell rheology.

\par
\begin{figure}
\centering
\begin{tabular}{cc}
\includegraphics[width=0.45\textwidth]{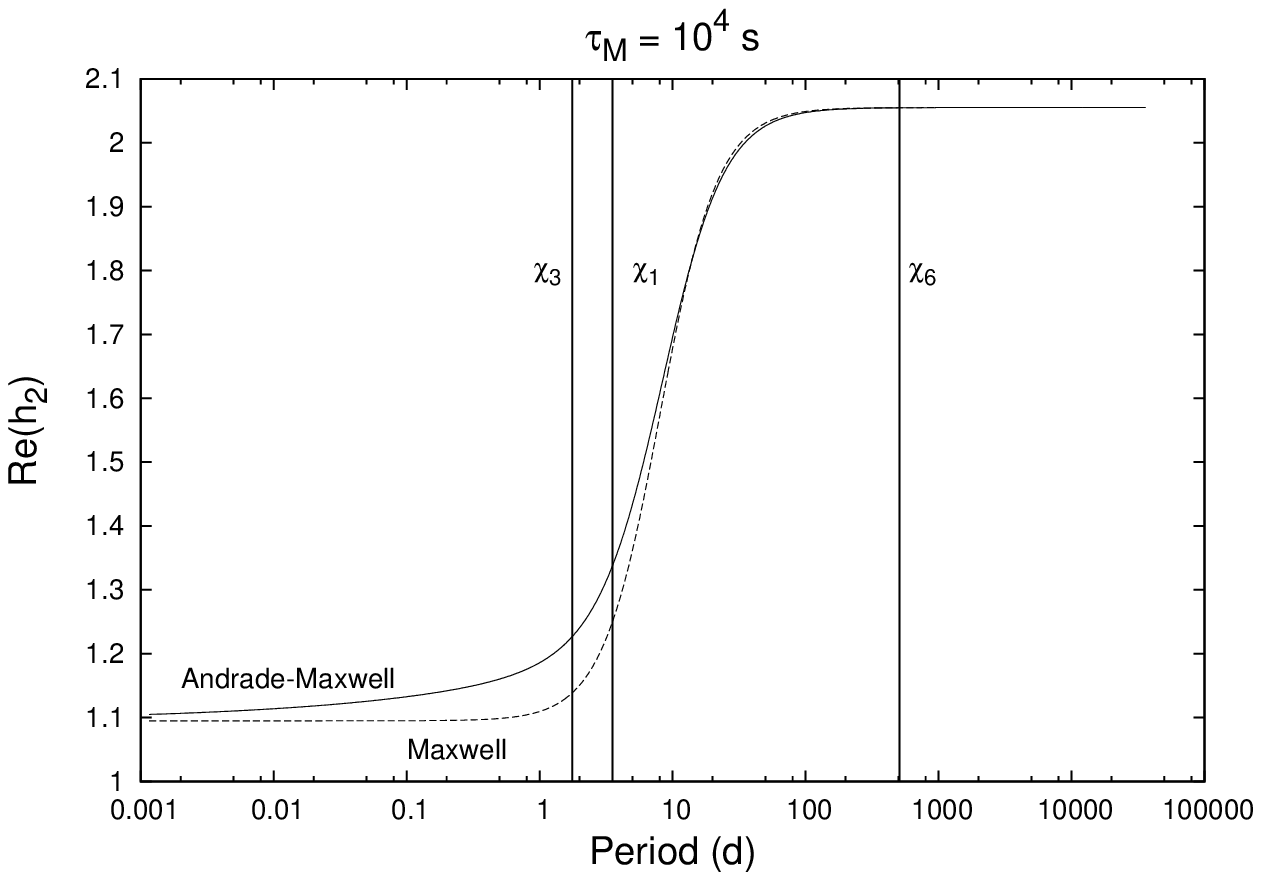} & \includegraphics[width=0.45\textwidth]{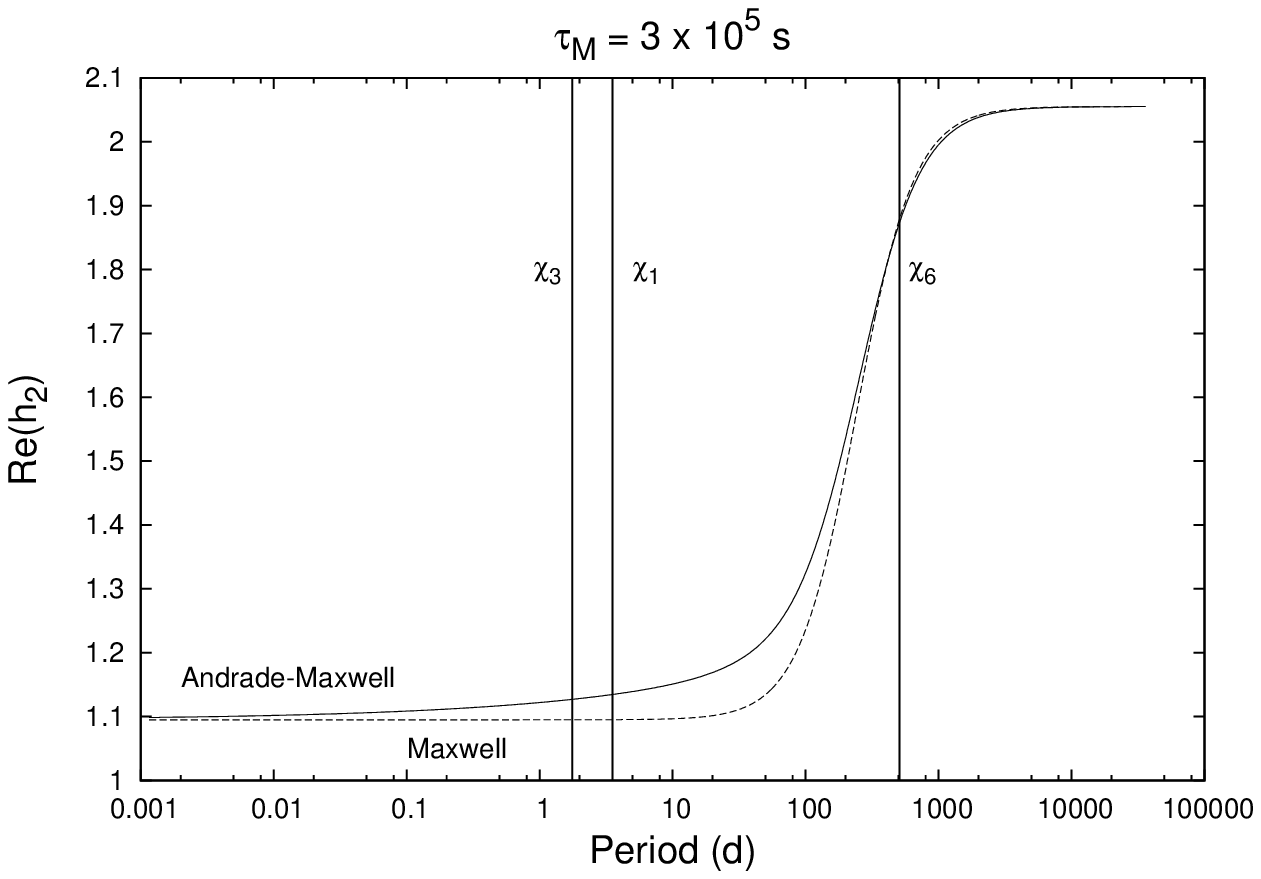} \\
\includegraphics[width=0.45\textwidth]{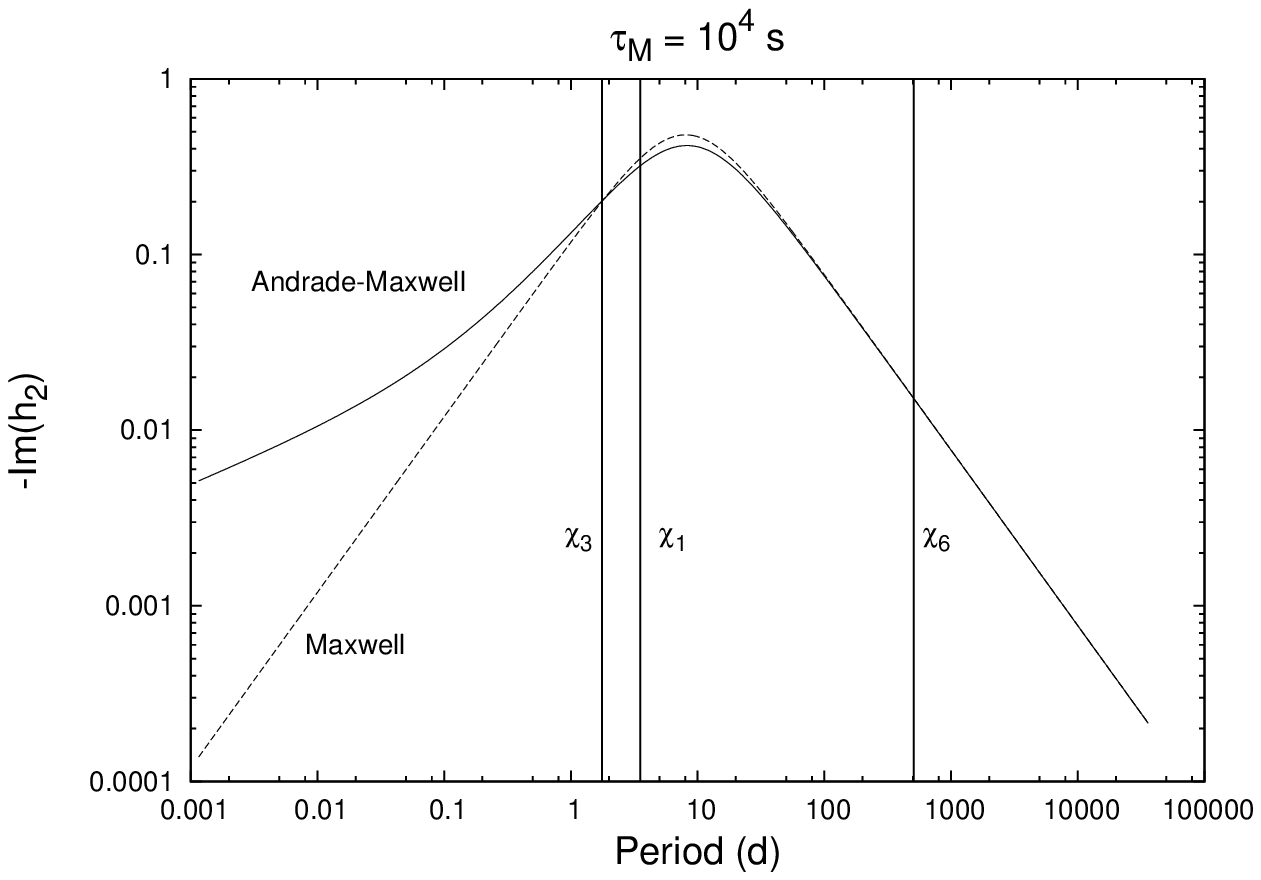} & \includegraphics[width=0.45\textwidth]{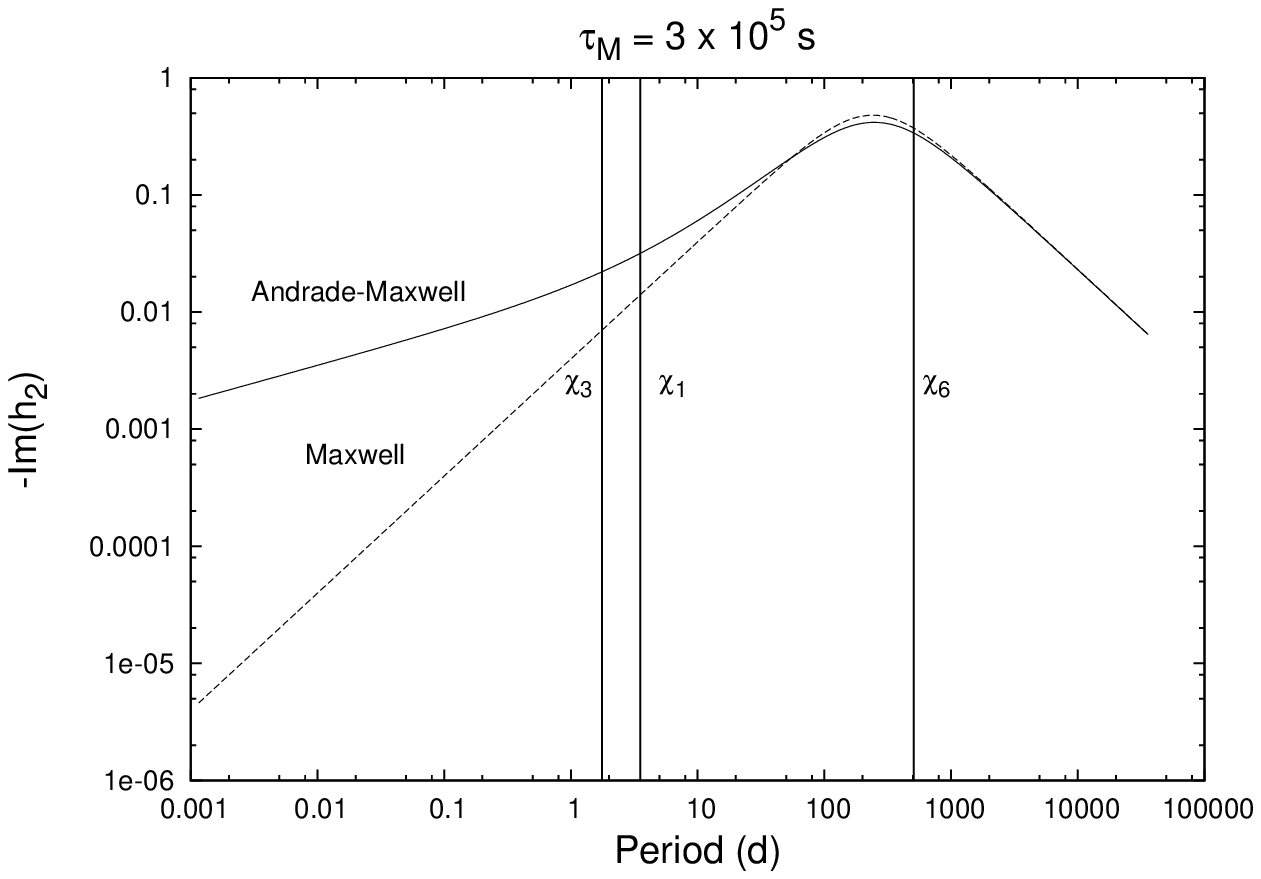}
\end{tabular}
\caption[Dependency of $h_2^*$ on the tidal frequency.]{Evolution of the tidal Love number $h_2^*$ with respect to the tidal frequency. The 3 vertical lines
locate the frequencies $\chi_1$, $\chi_3$ and $\chi_6$, present in the forced rotational dynamics of Europa. We can see a peak of dissipation in the 
imaginary part, that occurs at the period $2\pi(1+A_2)\tau_M$ for the Maxwell rheology.\label{fig:tidalmodels}}
\end{figure}

 The Fig.~\ref{fig:tidalmodels} and Tab.~\ref{tab:retidalmodels} \& \ref{tab:mimtidalmodels} illustrate the dependency of $h_2^*$ on the tidal frequency, for a thin shell that could be the one of 
Europa. For that, I take two different Maxwell times $\tau_M$, respectively of $10^4$ and $3\times10^5$ seconds, i.e. $0.116$ and $3.472$ days, that should be relevant for ice near its melting point
\citep[e.g.]{nm2009}, a rigidity $\mu = 2$ GPa as suggested by \citet{wzsl2006}, a density $\rho = 920$ $kg.m^{-3}$, an outer radius $R = 1.5608\times10^6$ m \citep{ntpm2007}, and a surface gravity 
$g=1.3147$ $m.s^{-2}$, consistent with $\mathcal{G}m = 3202.739$ $km^3.s^{-2}$, used for Europa in the JPL satellites ephemerides JUP230. I also used the mean values $h_2(\chi_0) = 2.05513$ and 
$1+k_2(\chi)-h_2(\chi) = 4.109\times10^{-4}$ for any tidal frequency. For the Maxwell-Andrade model, I set $\tau_A = \tau_M$ and
$\alpha=0.3$.

\par From \citep[Eq.~42]{b2015}, we would expect $k_2(\chi_1) = 0.6 (\rho_o/\bar{\rho}) h_2(\chi_1)$ for an infinitely rigid mantle. With $\rho_o = 1000$ kg.m$^{-3}$ and the mean density $\bar{\rho} = 3013$ kg.m$^-3$, 
that would correspond to $k_2(\chi_1)/h_2(\chi_1) = 0.199$, i.e. $\tau_M = 1.006\times10^4$ s, following the Maxwell model. This is consistent with the mean tidal viscosity of Enceladus of $2.4\times10^{13}$ Pa.s
found by \citet{e2018}, which would result in $\tau_M = 1.2\times10^4$ s for a rigidity $\mu = 2\times10^9$ Pa.

\par The main difference between the two tidal models appears in the imaginary part. As shown by the 
Tab.~\ref{tab:retidalmodels}, there is a limited number of frequencies present in the dynamics of the body, and there is no guarantee that a frequency that would be favourable to measure the tidal parameters 
would actually be present in the dynamics. In these tables, the frequencies $\chi_1$ to $\chi_6$ have been calculated from the main terms of the quasiperiodic decomposition of the orbital elements of Europa given 
by \citet{ldv2006}, i.e. $n=646.24512023768$ $rad/y$, $\dot{\varpi}=-4.71423211252$ $rad/y$, and $\dot{\ascnode}=-0.20790263943$ $rad/y$. A more complete dynamics would involve more frequencies 
(see Sec.~\ref{sec:orbital}), but their number would be limited anyway.

\begin{table}
\centering
\caption[Values of $\mathfrak{R}(h_2)$ at the main orbital frequencies]{Values of $\mathfrak{R}(h_2)$ at the main orbital frequencies.\label{tab:retidalmodels}}
\begin{tabular}{l|r|cc|cc}
\hline
Frequencies & Periods (d) & \multicolumn{2}{c}{$\mathfrak{R}(h_2)(\chi)$, $\tau_M = 10^4$ s} & \multicolumn{2}{c}{$\mathfrak{R}(h_2)(\chi)$, $\tau_M = 3\times 10^5$ s} \\
            &             & Maxwell              & Andrade-Maxwell & Maxwell              & Andrade-Maxwell \\
\hline
$\chi_1=\left|n-\dot{\varpi}\right|$                 &    $3.52546$ & $1.24962$ & $1.33733$ & $1.09519$ & $1.13463$ \\
$\chi_2=\left|n-\dot{\ascnode}\right|$               &    $3.55005$ & $1.25143$ & $1.33895$ & $1.09520$ & $1.13472$ \\
\hline
$\chi_3=2\left|n-\dot{\varpi}\right|$                &    $1.76273$ & $1.13896$ & $1.22703$ & $1.09504$ & $1.12706$ \\
$\chi_4=2\left|n-\dot{\ascnode}\right|$              &    $1.77502$ & $1.13954$ & $1.22773$ & $1.09504$ & $1.12713$ \\
$\chi_5=\left|2n-\dot{\varpi}-\dot{\ascnode}\right|$ &    $1.76885$ & $1.13925$ & $1.22738$ & $1.09504$ & $1.12710$ \\
\hline
$\chi_6=\left|\dot{\varpi}-\dot{\ascnode}\right|$    & $-509.26890$ & $2.05489$ & $2.05477$ & $1.87899$ & $1.87341$ \\
\hline
\end{tabular}
\end{table}

\begin{table}
\centering
\caption[Values of $-\mathfrak{I}(h_2)$ at the main orbital frequencies]{Values of $-\mathfrak{I}(h_2)$ at the main orbital frequencies.\label{tab:mimtidalmodels}}
\begin{tabular}{l|r|cc|cc}
\hline
Frequencies & Periods (d) & \multicolumn{2}{c}{$-\mathfrak{I}(h_2)(\chi)$, $\tau_M = 10^4$ s} & \multicolumn{2}{c}{$-\mathfrak{I}(h_2)(\chi)$, $\tau_M = 3\times 10^5$ s} \\
            &             & Maxwell              & Andrade-Maxwell & Maxwell              & Andrade-Maxwell \\
\hline
$\chi_1=\left|n-\dot{\varpi}\right|$                 &    $3.52546$ & $0.35293$ & $0.31983$ & $0.01402$ & $0.03174$ \\
$\chi_2=\left|n-\dot{\ascnode}\right|$               &    $3.55005$ & $0.35459$ & $0.32107$ & $0.01412$ & $0.03186$ \\
\hline
$\chi_3=2\left|n-\dot{\varpi}\right|$                &    $1.76273$ & $0.20071$ & $0.20213$ & $0.00701$ & $0.02209$ \\
$\chi_4=2\left|n-\dot{\ascnode}\right|$              &    $1.77502$ & $0.20198$ & $0.20315$ & $0.00706$ & $0.02217$ \\
$\chi_5=\left|2n-\dot{\varpi}-\dot{\ascnode}\right|$ &    $1.76885$ & $0.20134$ & $0.20264$ & $0.00704$ & $0.02213$ \\
\hline
$\chi_6=\left|\dot{\varpi}-\dot{\ascnode}\right|$    & $-509.26890$ & $0.01517$ & $0.01510$ & $0.37161$ & $0.33778$ \\
\hline
\end{tabular}
\end{table}

\section{The observable quantities} \label{sec:libra}

\par This section aims at studying the influence of the viscoelasticity of the crust on the observable rotational quantities. I derive these quantities
as if the rotation of a viscoelastic shell could be measured. Comparing the results with the ones of a rigid shell shows the influence of the elasticity.
These quantities are the longitudinal librations and the obliquity, for which I first propose a new way to see them, before deriving their formulae.

\subsection{Libration and obliquity}

\par The longitudinal librations are usually defined as the periodic oscillations at the orbital frequency of the spin rate around its mean value, that is the orbital frequency $n$.
Under this definition, they are named \emph{physical librations}, and their amplitude is proportional to the eccentricity $e$, which is well-known, and to the triaxiality \citep[e.g.]{md1999}. 
For rigid bodies, measuring the physical libration gives a direct measurement of the quantity $(I_{22}-I_{11})/I_{33}$, and this interpretation can be extended to bodies composed of a 
rigid layer enshrouding a fluid core \citep{n2012}. If the outer layer has some elasticity, then a repelling torque tends to counterbalance the orbital forcing of the parent body \citep{gm2010},
diminishing the amplitude of libration. Physical librations of synchronous bodies have been measured for the Moon \citep{k1967}, Phobos \citep{b1972}, Epimetheus \citep{ttb2009},
 Mimas \citep{trlcrrn2014}, Enceladus \citep{tttbjlhp2016}, Pandora, and Prometheus \citep{lcmnprrt2016}.

\par The obliquity $\varepsilon$ is defined as the instantaneous angle between the angular momentum and the orbital plane. It can be shown \citep[e.g.]{wh2004,n2010} that for a despun rotation, 2 or 4 positions,
named \emph{Cassini States}, are possible, and that the number of these positions depends on the precession rate of the orbital node. 4 Cassini States exist for most of the satellites of the giant planets,
while only 2 for the Moon. Some of these states are unstable, the Moon is actually in the Cassini State 2, and the satellites of the giant planets are expected to lie in the Cassini State 1. This is 
a stable state corresponding to a small obliquity. For a rigid body whose orbit precesses uniformly with a inclination $I$ and a precession rate $\dot{\ascnode}$ over a reference plane, usually
named \emph{Laplace Plane}, we have 

\begin{equation}
\label{eq:cs1rigid}
\varepsilon \approx -\frac{\sin I}{\alpha/\dot{\ascnode}+\cos I},
\end{equation}
with 

\begin{equation}
\label{eq:alpharigid}
\alpha=\frac{3}{2}n\frac{I_{33}-I_{11}}{I_{33}},
\end{equation}
so, measuring the obliquity is measuring the quantity $(I_{33}-I_{11})/I_{33}$. This formula (Eq.\ref{eq:cs1rigid}) is invalid when a fluid core is involved \citep{n2012}, and of course with a more complex
structure \citep[e.g.]{bvyk2011}.

\subsection{The physical meaning of the librations}

\par When the rotation of a planetary satellite is despun, its rotation is a forced motion directly related to the tidal torque exerted by its parent planet. The variations of this motion, 
i.e. the librations, are due to variations of this torque. In \citep{n2010} I defined the \emph{tidal librations}, with respect to the satellite-planet direction. These librations
can be decomposed into the \emph{tidal longitudinal librations} and the \emph{tidal latitudinal librations}. 

\par The tidal longitudinal librations differ from the physical librations by the optical librations. It is actually a good approximation to consider that the long axis of the satellite 
points to the empty focus of its orbit, while its parent planet is located at the other focus. If we neglect any physical response of the interior of the satellite, then we have an
optical libration of amplitude twice the eccentricity. Denoting $\phi$ the optical libration, $\gamma$ the physical one, and $\psi$ the tidal one, they are bound by the following relation:

\begin{equation}
\label{eq:optical}
\phi = 2e\sin nt = \gamma-\psi.
\end{equation}

\par The amplitude of the tidal libration $\psi$ is dominated by the eccentricity alone, making it unpopular to get the internal structure. However, it helps to give a picture of
the phenomenon, since its geometry is based on the tidal torque. If we neglect the influence of the triaxiality, then the amplitude of the libration is $2e$. The triaxiality tends to 
align the long axis of the satellite with the direction of the planet to reduce the tidal torque, so it reduces the amplitude of libration. If the satellite has some elasticity, then 
its shape has a dependency on the tidal torque. That will result in a displacement of the long axis in the reference frame of the satellite, reducing the tidal torque, thus 
needing less reorientation. So, as pointed out by previous studies, it counterbalances the effect of the triaxiality, lowering the amplitude of the physical librations but raising the one of
the tidal librations. 

\par The obliquity is the amplitude of the tidal latitudinal librations. The elasticity acts the same way, in counterbalancing the latitudinal reorientation of the satellite by altering its
shape, and so the elasticity raises the obliquity.

\subsection{The longitudinal librations\label{sec:longilibr}}

\par Similar calculations of the longitudinal librations can be found in \citep{rcwk2010,vbt2013,rrc2014,mfd2016}, their differences being discussed in \citep{n2017}.

I assume that the planar motion is a good enough approximation, which induces that the rotation of the crust is ruled only by the quantities $I_{22}-I_{11}$, $I_{12}$
and $I_{33}$. Moreover, the planetary perturbations are considered, which means that the mean longitude of the body in its orbital motion around its parent planet 
should be written as

\begin{equation}
\label{eq:meanlongi}
\lambda(t) = \lambda_0+nt+\sum_i H_i\sin(\omega_i t+\alpha_i) = \lambda^*(t)+\sum_i H_i\sin(\omega_i t+\alpha_i).
\end{equation}

The equation ruling the longitudinal motion can be written in equating two expressions for the z-component of the gravitational torque $\vec{\Gamma}$, i.e. $\Gamma_3$. The Eq.~\ref{eq:Gamma3}
becomes, with $z=0$:

\begin{equation}
\label{eq:Gamma32}
\Gamma_3 = 3n^2\left(\frac{a}{r}\right)^3\left(\left(I_{22}-I_{11}\right)xy+I_{12}\left(x^2-y^2\right)\right),
\end{equation}
which yields, in using the Eqs~\ref{eq:asr},\ref{eq:cosf},\ref{eq:sinf}:

\begin{eqnarray}
\Gamma_3 & = & \frac{48}{5}\pi\frac{\rho}{g}n^4R^2X_3\left[-h_2(\chi_0)\sigma+2e\left((h_2(\chi_0)-h_2(\chi_1))\sin(\lambda-\varpi)+\frac{h_2}{Q}(\chi_1)\cos(\lambda-\varpi)\right)\right.  \nonumber \\
         &   & \left.+\sum_i\left(\left(h_2(\chi_0)-h_2(\omega_i)\right)H_i\sin(\omega_it+\alpha_i)+\left(\frac{h_2}{Q}\right)(\omega_i)H_i\cos(\omega_it+\alpha_i)\right)+6e^2\left(\frac{h_2}{Q}\right)(\chi_1)\right]. \label{eq:Gamma3longi1}
\end{eqnarray}

Moreover, we have from $\Gamma_3 = d\vec{G}/dt\cdot\hat{f_3} = d(I_{33}\dot{p})/dt$:

\begin{equation}
\label{eq:Gamma3longi2}
\Gamma_3 = \left(\frac{8}{3}\pi\rho X_4+\frac{16}{9}\pi\frac{n^2R^2}{g}\rho X_3 h_2(\chi_0)\right)\ddot{\sigma} \approx \mathcal{C}\ddot{\sigma}.
\end{equation}
In these equations, $\sigma$ is the argument of the resonant synchronous rotation, defined as

\begin{equation}
\label{eq:sigma}
\sigma = p-\lambda+\pi,
\end{equation}
and the expressions are expanded up to the degree 1 in the eccentricity, except for the constant term, which induces a shift of the mean value of the resonant argument.
This shift is due to the dissipation in the body, which delays its response \citep{wbyrd2001,n2017}. Equating the expressions (\ref{eq:Gamma3longi1}) and (\ref{eq:Gamma3longi2})
gives

\begin{equation}
\label{eq:edolongitude}
\ddot{\sigma}+\omega_0^2\sigma = \kappa_0+\kappa_1\sin(\lambda^*-\varpi)+\kappa_2\cos(\lambda^*-\varpi)+\sum_i\left(\kappa_1^i\sin(\omega_it+\alpha_i)+\kappa_2^i\cos(\omega_it+\alpha_i)\right),
\end{equation}
and 

\begin{equation}
\label{eq:sigmasolved}
\sigma(t) = \sigma_0+\sigma_1\sin(\lambda^*-\varpi+\phi_1)+\sum_i\sigma_i\sin(\omega_i t+\phi_i),
\end{equation}
with

\begin{eqnarray}
\omega_0^2 & = & \frac{54n^4R^2X_3h_2(\chi_0)}{15gX_4+10n^2R^2X_3h_2(\chi_0)}, \label{eq:omega02} \\
\sigma_0 & = & 6e^2\frac{(h_2/Q_h)(\chi_1)}{h_2(\chi_0)}, \label{eq:sigma0} \\
\sigma_1 & = & \frac{\frac{96}{5}\frac{n^4R^2}{g}X_3e\sqrt{\left(h_2(\chi_0)-h_2(\chi_1)\right)^2+\left(h_2/Q_h\right)(\chi_1)^2}}{\left(\frac{8}{3}X_4+\frac{16}{9}\frac{n^2R^2}{g}X_3h_2(\chi_0)\right)\left(\omega_0^2-(n-\dot{\varpi})^2\right)}, \label{eq:sigma1} \\
\phi_1 & = & \arctan\left(\frac{\left(h_2/Q_h\right)(\chi_1)}{h_2(\chi_0)-h_2(\chi_1)}\right), \label{eq:phi1} \\
\sigma_i & = &  \frac{\frac{48}{5}\frac{n^4R^2}{g}X_3H_i\sqrt{\left(h_2(\chi_0)-h_2(\omega_i)\right)^2+\left(h_2/Q_h\right)(\omega_i)^2}}{\left(\frac{8}{3}X_4+\frac{16}{9}\frac{n^2R^2}{g}X_3h_2(\chi_0)\right)\left(\omega_0^2-\omega_i^2\right)}, \label{eq:sigmai} \\
\phi_i & = & \arctan\left(\frac{\left(h_2/Q_h\right)(\omega_i)}{h_2(\chi_0)-h_2(\omega_i)}\right)-\alpha_i. \label{eq:phii}
\end{eqnarray}

\subsection{The obliquity}

\par Here we consider a circular orbit, with a constant inclination $I$, the ascending node $\ascnode$ precessing uniformly at the constant rate $\dot{\ascnode}$.
The polar motion is neglected as well, i.e. $J=0$.
The coordinates of the unit vector pointing to the satellite are, in the inertial reference frame:

\begin{eqnarray}
x_i & = & -(\cos\ascnode\cos(\lambda-\ascnode)-\cos I\sin\ascnode\sin(\lambda-\ascnode)), \label{eq:xioblik} \\
y_i & = & -(\sin\ascnode\cos(\lambda-\ascnode)+\cos I\cos\ascnode\sin(\lambda-\ascnode)), \label{eq:yioblik} \\
z_i & = & -\sin I\sin(\lambda-\ascnode), \label{eq:zioblik}
\end{eqnarray}
while the Hamiltonian (\ref{eq:energyrot}) becomes

\begin{equation}
\label{eq:Hamiltonoblik}
\mathcal{H} = \frac{nP^2}{2}-\frac{3}{2}n\left(\gamma_1\left(x^2+y^2\right)+\gamma_2xz+\gamma_4\left(x^2-y^2\right)\right),
\end{equation}
in neglecting the second degree in the obliquity, i.e. $\varepsilon^2$,
with

\begin{eqnarray}
\gamma_1 & = & \frac{8}{3}\pi n^2R^2X_3\frac{\rho}{g\mathcal{C}}h_2(\chi_0), \label{eq:gamma1oblik} \\
\gamma_2 & = & \frac{32}{5}\pi n^2R^2X_3\frac{\rho}{g\mathcal{C}}\left(\sin K_0-\sin I\right)\left(h_2(\chi_2)\sin(\lambda-\Omega)-\left(\frac{h_2}{Q}\right)(\chi_2)\cos(\lambda-\Omega)\right), \label{eq:gamma2oblik} \\
\gamma_4 & = & \frac{8}{5}\pi n^2R^2X_3\frac{\rho}{g\mathcal{C}}h_2(\chi_0). \label{eq:gamma4oblik}
\end{eqnarray}
In this formulae, the quantity $K$, which is defined as the angle between the angular momentum and the third axis of the inertial reference frame $\hat{e}_3$ (cf. Sect.~\ref{sec:potential}), 
is denoted $K_0$ since it is considered as frozen in the coefficients of the Hamiltonian, and as such will not be affected by the derivations. This issue is discussed in Sect.~\ref{sec:frozen}.
Once I have substituted the expressions for $\gamma_1$, $\gamma_2$ and $\gamma_4$ into the Hamiltonian, I

\begin{enumerate}
\item express the slow arguments $\sigma = l+g+h-\lambda+\pi$ (synchronous spin-orbit resonance) and $\rho' = \ascnode+h$ (Third Cassini Law). This is a time-varying transformation since 
$\lambda$ and $\ascnode$ depend linearly on the time, their derivatives being $n$ and $\dot{\ascnode}$, respectively. For this reason, $\dot{\ascnode}R-nP$ must be added to the Hamiltonian,
\item average over the circulating angles $\lambda$ and $\ascnode$,
\item express the strict spin-orbit resonance as $\sigma = 0$ and $P = 1$.
\end{enumerate}

This way, the averaged Hamiltonian of the problem (Eq.~\ref{eq:Hamiltonoblik}) reads

\begin{eqnarray}
<\mathcal{H}> & = & \frac{24}{5}\pi n^3R^2X_3\frac{\rho}{g\mathcal{C}}\sin K \left(-\frac{4}{3}h_2(\chi_0)\cos\rho'\sin I+
h_2(\chi_2)\left(\sin I-\sin K_0\right)\cos\rho'+\left(\frac{h_2}{Q}\right)(\chi_2)\left(\sin K_0-\sin I\right)\sin\rho'\right) \nonumber \\
& + & \sin^2K\left(\frac{\dot{\ascnode}}{2}+\frac{16}{5}\pi n^3R^2X_3h_2(\chi_0)\frac{\rho}{g\mathcal{C}}\right), \label{eq:Hamiltonoblik2}
\end{eqnarray}
its variables being $R = 1-\cos K$ and $\rho'$. $R$ only appears as $\sin K$, the $R$ appearing explicitly in the Hamiltonian being the mean radius. At the equilibrium, the derivatives of the 
Hamiltonian should be null. The derivations of $<\mathcal{H}>$ with respect to $\rho'$ and $\sin K$ give respectively:

\begin{equation}
\label{eq:dHrho}
3\sin K_0\left(h_2(\chi_2)\sin\rho'+\left(\frac{h_2}{Q}\right)(\chi_2)\cos\rho'\right) = \sin I\left(-4h_2(\chi_0)+3h_2(\chi_2)\sin\rho'+\left(\frac{h_2}{Q}\right)(\chi_2)\cos\rho'\right),
\end{equation}
and

\begin{eqnarray}
\sin K\left(\dot{\ascnode}+\frac{32}{5}\pi n^3R^2X_3\frac{\rho}{g\mathcal{C}}h_2(\chi_0)\right) & &  \label{eq:dHsK} \\
+\frac{24}{5}\pi n^3R^2X_3\frac{\rho}{g\mathcal{C}}\sin K_0\left(\left(\frac{h_2}{Q}\right)(\chi_2)\sin\rho'-h_2(\chi_2)\cos\rho'\right) & = & \frac{24}{5}\pi n^3R^2X_3\frac{\rho}{g\mathcal{C}}\sin I\left(\frac{4}{3}h_2(\chi_0)\cos\rho'+\left(\frac{h_2}{Q}\right)(\chi_2)\sin\rho'-h_2(\chi_2)\cos\rho'\right). \nonumber
\end{eqnarray}

\par From $K_0 = K = \varepsilon+I$ and the Eq.~\ref{eq:dHrho} I get

\begin{equation}
\label{eq:rhoequi}
\tan\rho' = -\frac{3\left(\frac{h_2}{Q}\right)(\chi_2)\varepsilon}{4h_2(\chi_0)\sin I+3h_2(\chi_2)\varepsilon},
\end{equation}
which means that the dissipation at the frequency $n-\dot{\ascnode}$ induces a small time lag of the orientation of the Cassini Plane, i.e. a small violation of the expected coplanarity
between the spin axis and the normals to the ecliptic and orbit planes. Such a deviation has been predicted for Mercury \citep{pmhs2014,byrv2017}.

\par This is anyway a very small deviation, which permits to neglect it in Eq.~\ref{eq:dHsK} for deriving the equilibrium obliquity, which is then:

\begin{equation}
\label{eq:obliqequi}
\varepsilon = \frac{-\tan I}{1+\frac{n}{\dot{\ascnode}}\pi\frac{n^2R^2}{5g\mathcal{C}}\rho X_3\left(32 h_2(\chi_0)-24 h_2(\chi_2)\right)}.
\end{equation}
We can see the influence of two Love numbers, which are $h_2(\chi_0)$ and $h_2(\chi_2)$. $h_2(\chi_0)$ refers to the permanent component of the inertia tensor, and neglecting $h_2(\chi_2)$
renders the classical formula (\ref{eq:cs1rigid}). $h_2(\chi_2)$ comes from the product of inertia $I_{13}$ and renders the oscillating alteration of the shape of the satellite, which counterbalances
its reorientation and raises the amplitude of the latitudinal libration, i.e. the obliquity $\varepsilon$. An alternative modelisation of this effect can be found in \citep{byv2016}.

\subsection{Frozen vs. hydrostatic inertia\label{sec:frozen}}

\par An issue is whether the moments of inertia should be considered as constants or not when deriving the equations of the rotational dynamics. I have dedicated a full paper on this problem
\citep{n2017} and showed that this yields different results for the longitudinal librations. This is why I have proposed a mixed model in \textit{Ibid.} in which I consider that part of the 
tensor of inertia is frozen, the other part being to be derived. 

\par Actually many authors consider that the moments of inertia are frozen, even if they have been estimated from the theory of the hydrostatic equilibrium. This is the choice I make here, 
which is consistent with the use of a numerical iterative algorithm for computing the rotational response of the crust (Sect.~\ref{sec:numerical}). Deriving the moments of 
inertia in deriving the equations of motion would have changed the Eq.~\ref{eq:obliqequi} into

\begin{equation}
\label{eq:obliqequiother}
\varepsilon = \frac{-\tan I}{1+\frac{n}{\dot{\ascnode}}\pi\frac{n^2R^2}{g\mathcal{C}}\rho X_3\left(16 h_2(\chi_0)-\frac{48}{5} h_2(\chi_2)\right)}.
\end{equation}

\section{Application to the crust of Europa}

I here present the application of the algorithm to the rotation of a viscoelastic shell, which would be the one of Europa. For that, I first create a set a model-shells based on the
currrently known physical characteristics of Europa (Sect.~\ref{sec:interior}), on which I apply the gravitational torque of Jupiter. This torque considers the realistic orbital motion of Europa around Jupiter,
which requires to use the planetary perturbations of the other satellites given by ephemerides (Sect.~\ref{sec:orbital}).

\subsection{Interior model \label{sec:interior}}

\par For creating the model-shells, I first create model-Europas, on which I impose the following constraints:

\begin{itemize}

\item its shape is hydrostatic,

\item $\mathcal{G}m = 3,202.72$ km$^3$.s$^{-2}$ \citep{asjlms1998}

\item $C_{22} = 1.315\times10^{-4}\pm2.5\times10^{-6}$ \citep{asjlms1998},

\item the along orbit equatorial radius $b$ is equal to $1,560.3$ km. \citep{ntpm2007}

\end{itemize}

\par As in \citep{bv2010}, Europa is assumed to be composed of 4 homogeneous layers, which are:

\begin{enumerate}

\item an icy shell,

\item a global water ocean,

\item a rock mantle,

\item a core.

\end{enumerate}

To derive the model-Europas, I start from the 6 following inputs: the mean depths and densities of the 3 outer layers, the mean radius $R$ of Europa being set to $1,560.8$ km. Then

\begin{enumerate}

\item the core radius, its density, and the resulting polar momentum of inertia $\mathcal{C}$ are deduced,

\item the secular Love number $h_2(\chi_0)$ is obtained from 

\begin{equation}
\label{eq:h2hydro}
h_2(\chi_0) = \frac{5}{1+\left(\frac{5}{2}-\frac{15}{4}\frac{C}{mR^2}\right)^2},
\end{equation}

\item the 2 missing radii $a$ and $c$ are derived from the Eqs.(\ref{eq:chydro}) to (\ref{eq:qhydro}).

\item The knowledge of the 3 surface radii $a$, $b$, and $c$ give the flattening $\alpha=(a+b-2c)/(a+b)$ at the radius $R$. The radii of the three boundary layers are given in propagating the 
Clairaut equation, which gives the flattening of the equipotential surface at any radius $r$:

\begin{eqnarray}
\frac{\textrm{d}^2\alpha}{\textrm{d}r^2}+\frac{6}{r}\frac{\rho}{\bar{\rho}}-\frac{6}{r^2}\left(1-\frac{\rho}{\bar{\rho}}\right) & = & 0, \label{eq:clairauteq} \\
\frac{\textrm{d}\alpha}{\textrm{d}r}(R) & = & \frac{1}{R}\left[\frac{25}{4}q-2\alpha(R)\right], \label{eq:clairautci1} \\
\alpha(R) & = & \frac{a+b-2c}{a+b}, \label{eq:clairautci2}
\end{eqnarray}
where $\rho$ is the density at the radius $r$, and $\bar{\rho}$ is the mean density of mass enshrouded by the equipotential surface at the radius $r$. The initial condition (\ref{eq:clairautci1}) is not the classical one given in the Clairaut theory where $25/4$ should be replaced by $5/2$, since the tides are not considered in the original theory. The information on $\alpha$ is supplemented by the knowledge of $\beta = (a-b)/a$. At any radius, we have $\beta = 6\alpha/5$ \citep{vrkdr2008}.

\item The Stokes coefficient $C_{22}$ is calculated from the internal structure, and the model is kept only if $C_{22}$ complies with the measurements by Galileo, i.e. $C_{22} = 1.315\times10^{-4}\pm2.5\times10^{-6}$ \citep{asjlms1998}.

\end{enumerate}

\begin{table}
\centering
\caption{Input parameters for the computation of the model-Europas. Most of these parameters are consistent with the ones chosen by \citet[Tab.~3]{bv2010}. The shell thickness $\eth$ is
limited to 40 km since \citet{ntpm2007} suggest 35 km to be an upper limit.\label{tab:intinput}}
\begin{tabular}{lc}
\hline
Shell & \\
\hline
Density $\rho_s$ & 800 - 1200 kg/m$^3$ \\
\hline
Thickness $\eth$ & 3-40 km \\
\hline\hline
Ocean & \\
\hline
Density $\rho_o$ & $\rho_s$ - 1200 kg/m$^3$ \\
\hline
Thickness & 10 - 175 km \\
\hline\hline
Mantle & \\
\hline
Density & 3000 - 3900 kg/m$^3$ \\
\hline
Thickness & 300 - 1400 km \\
\hline
\end{tabular}
\end{table}

\par This way, I get 33,935 model-Europas, for which $h_2(\chi_0) = 2.05513$ with a standard deviation of 0.01152, and an ocean thickness of 101.329 km, with a standard deviation of 25.5785 km. 

\subsection{Orbital model \label{sec:orbital}}

\par The orbital dynamics of Europa is complexified by the presence of the three other Galilean satellites, i.e. Io, Ganymede and Callisto, which are massive enough to significantly
perturb it. In particular, Io, Europa and Ganymede are trapped into a 3-body $1\!:\!2\!:\!4$ mean-motion resonance, known as \textit{Laplacian resonance}, and Europa is close to two 
$2\!:\!1$ mean-motion resonances, with Io and Ganymede.
Fortunately, the time evolution of the orbital osculating elements of Europa can be given under the form of converging quasi-periodic series, i.e. as sums of sinusoides, as in \citet{ldv2006}.
This is particularly convenient here, since the contributions of the different tidal frequencies will be naturally distinguished.

\par For that, fundamental frequencies ruling the motion of Europa are identified, see Tab.~\ref{tab:frekfon}. The time evolution of the jovicentric orbital osculating elements of Europa
is considered to be fully describable with these frequencies. In practice, \citet{ldv2006} noted some very small discrepancy from this rule in their conservative model, I neglect it since
such an accuracy is not needed in this study. Considering the tidal dissipation in the orbital motion of Europa would induce a secular change of these frequencies, but this is a very long-term
process, which can be safely neglected here.

\begin{table}
\centering
\caption[The fundamental frequencies ruling the orbital motion of Europa]{The fundamental frequencies ruling the orbital motion of Europa, from \citet{ldv2006}. The phase origin
is JD 2,433,282.5. Their denomination is inspired from \citet{ldv2006}, the indices i referring to the Galilean satellite J-i, in which the identified sinusoidal contribution
has the most influence. $\nu = \lambda_1-2\lambda_2 = \lambda_2-2\lambda_3$ is the signature of the proximity of the $2\!:\!1$ MMR with Io and Ganymede, and $\psi$ is the
libration of the argument of the Laplacian resonance.\label{tab:frekfon}}
\begin{tabular}{rrrrr}
\hline
Frequency & Phase & Period & Period & Identification \\
(rad/y) & (deg) & (d) & (y) & \\
\hline
  $646.2451202376758$ & $338.598517$ &       $3.5512$ &   $0.0097$ & $\lambda_2$ \\
  $4.71423212969175$  & $125.663720$ &     $486.8096$ &   $1.3328$ & $\nu$ \\
  $-0.20790263942625$ &  $60.029150$ &  $11,038.5007$ &  $30.2218$ & $\ascnode_2$ \\
  $0.2476297953075$   & $157.191935$ &   $9,267.5981$ &  $25.3733$ & $\varpi_2$ \\
  $0.0464868759825$   & $121.789605$ &  $49,367.3404$ & $135.1604$ & $\varpi_3$ \\
  $0.01171177740975$  & $319.917125$ & $195,950.9091$ & $536.4844$ & $\varpi_4$ \\
  $0.97311853791375$  &  $25.288760$ &   $2,358.3288$ &   $6.4568$ & $\varpi_1$ \\
  $0.52967961432225$  & $318.603037$ &   $4,332.6822$ &  $11.8622$ & $\lambda_{\sun}$ \\
  $-0.04562449918275$ & $191.189501$ &  $50,300.4630$ & $137.7152$ & $\ascnode_3$ \\
$1,297.2044725279754$ &  $82.861918$ &       $1.7691$ &   $0.0048$ & $\lambda_1$ \\
 $1.114249427448$     & $109.319546$ &   $2,059.6227$ &   $5.6389$ & $\psi$ \\
\hline
\end{tabular}
\end{table}

\par To express the orbital elements, I use the tables 7 to 10 of \citep{ldv2006}, with a cut-off at the amplitude of 120 km. These elements are

\begin{itemize}

\item the semimajor axis $a$, which is set to a constant, i.e. $a = 671,261.171$ km,

\item the mean longitude $\lambda(t)$ (Tab.~\ref{tab:lambdavalery}),

\item $z(t) = e(t)\exp(\imath\varpi(t))$, $e$ being the eccentricity of Europa, and $\varpi$ the longitude of its pericentre (Tab.~\ref{tab:zvalery}),

\item $\zeta(t) = \sin\left(I/2\right)(t)\exp(\imath\ascnode(t))$, $I$ being the inclination of Europa, and $\ascnode$ the longitude of its ascending node (Tab.~\ref{tab:zetavalery}).

\end{itemize}
$\lambda$, $z$ and $\zeta$ are given in km in \citep{ldv2006} since they are multiplied by the mean value of the semimajor axis. I here present them without this multiplication, i.e. in radians. In all of these tables, 
the time origin is the same as in \citep{ldv2006}, i.e. JD 2,433,282.5 = 1950/01/01, 0h UT.

\begin{table}
\centering
\caption{The mean longitude of Europa $\lambda(t)$, as a series expressed in sine. This series should be added to the slope $\lambda_0+nt$ with $\lambda_0 = 0.3735263437471362$.\label{tab:lambdavalery}}
\begin{tabular}{rrrrrrrrrrr}
Amplitude & $\lambda_2$ & $\nu$ & $\varpi_2$ & $\varpi_3$ & $\varpi_4$ & $\lambda_{\sun}$ & $\lambda_1$ & $\psi$  & Period (d) \\
\hline
 $8.576\times10^{-4}$ & -    & $1$  & -   & $1$ & -   & -   & -   & -   &   $482.0561$ \\
 $4.550\times10^{-4}$ & -    & $1$  & -   &   - & $1$ & -   & -   & -   &   $485.6032$ \\
 $3.249\times10^{-4}$ & $-1$ & -    & -   &   - & -   & -   & $1$ & -   &     $3.5255$ \\
$-3.074\times10^{-4}$ & -    & $1$  & $1$ & -   & -   & -   & -   & -   &   $462.5146$ \\
 $1.982\times10^{-4}$ & -    & -    & -   &   - & -   & -   & -   & $1$ & $2,059.6227$ \\
 $1.834\times10^{-4}$ & -    & -    & -   &   - & -   & $1$ & -   & -   & $4,332.6822$ \\
\hline
\end{tabular}
\end{table}

\begin{table}
\centering
\caption{$z(t) = e(t)\exp(\imath\varpi(t))$, expressed as a series in complex exponentials.\label{tab:zvalery}}
\begin{tabular}{rrrrrr}
Amplitude & $\lambda_2$ & $\nu$ & $\varpi_2$ & $\varpi_3$ & Period (d) \\
\hline
$-9.359\times10^{-3}$ & -   & $-1$ &  -   & -   &    $486.8096$ \\
 $2.989\times10^{-4}$ & $1$ & -    &  -   & -   &      $3.5512$ \\
 $2.139\times10^{-4}$ & -   & -    &  -   & $1$ & $49,367.3404$ \\
 $1.981\times10^{-4}$ & -   & -    &  $1$ & -   &  $9,267.5981$ \\
\hline
\end{tabular}
\end{table}

\begin{table}
\centering
\caption{$\zeta(t) = \sin\left(I/2\right)(t)\exp(\imath\ascnode(t))$, expressed as a series in complex exponentials.\label{tab:zetavalery}}
\begin{tabular}{rrrr}
Amplitude & $\ascnode_2$ & $\ascnode_3$ & Period (d) \\
\hline
 $4.040\times10^{-3}$ & $1$ & -   & $11,038.5007$ \\
 $2.200\times10^{-4}$ & -   & $1$ & $50,300.4630$ \\
\hline
\end{tabular}
\end{table}

\section{Numerical resolution}\label{sec:numerical}

\subsection{An iterative algorithm}

\par The numerical code I use to simulate the rotation of the viscoelastic crust uses as inputs:

\begin{itemize}

\item the crustal density $\rho_s$

\item the secular Love number $h_2(\chi_0)$

\item the crustal thickness $\eth$

\item 3 tidal parameters, i.e. the Maxwell time $\tau_M$, the Andrade time $\tau_A$, which is usually set equal to $\tau_M$, and the Andrade parameter $\alpha$.

\end{itemize}

\par In addition to these physical characteristics of the crust I consider, the code requires constants, which are gathered in Tab.~\ref{tab:constants}.

\begin{table}
\centering
\caption[Constants used in the numerical code]{Constants used in the numerical code. These constants are supplemented by the frequencies and phases of the fundamental modes of the ephemerides 
(Tab.~\ref{tab:frekfon}). The eccentricity and inclination are useful to derive preliminary time evolutions of the moments of inertia. \label{tab:constants}}
\begin{tabular}{|l|r|c|}
\hline
$\mathcal{G}M_p$ & $126,686,536.1$ & \citet{j2013} \\
$\mathcal{G}m$ & $3,202.72$ km$^3$.s$^{-2}$ & \citet{asjlms1998} \\
Mean radius $R$ & $1,560.8$ km & \citet{ntpm2007} \\
$e$ & $9.359\times10^{-3}$ & Tab.~\ref{tab:zvalery} \\
$I$ & $8.081\times10^{-3}$ rad & Tab.~\ref{tab:zetavalery} \\
$\mu$ & 2 GPa & \citet{wzsl2006} \\
\hline
\end{tabular}
\end{table}

Then the algorithm consists of the following steps:

\begin{enumerate}

\item Numerical integration of the equations (\ref{eq:eqhamilton}), derived from the Hamiltonian (\ref{eq:energyrot}). These equations use the tensor of inertia given by the formulae 
(\ref{eq:I11freq}) to (\ref{eq:I33freq}) in the elastic case, and (\ref{eq:I11qfreq}) to (\ref{eq:I33qfreq}) in the viscoelastic one. The numerical integrations are performed with the 
Adams-Bashforth-Moulton 10th order predictor corrector integrator \citep[e.g.]{hnw1993}. 
The initial conditions correspond to the $1\!:\!1$ spin-orbit resonance, which implies $P=1$, $p=\sigma_0+\lambda-\pi$, $r = \rho_0-\ascnode$, very small
numbers for $\xi$ and $\eta$, since the polar motion is expected to have a very small amplitude, and $R = 1-\cos\varepsilon$. $\sigma_0$, $\rho_0$ and $\varepsilon$ are derived from the formulae
(\ref{eq:sigma0}, \ref{eq:rhoequi}) and (\ref{eq:obliqequi}), respectively. New solutions for the moments of inertia are simultaneously computed using the formulae (\ref{eq:I11cart}) to (\ref{eq:I33cart}).
These new moments of inertia use a Love number, which does not depend on the tidal frequency.

\item Frequency analysis of the resulting moments of inertia and the 6 variables of the rotation. These 12 quantities are thus decomposed into quasi-periodic time series. The frequencies of the 
components of each of these series tell us whether the relevant component is forced by the variations of the gravitational torque of Jupiter, or is a free oscillation due to a departure from the 
exact spin-orbit resonance. These free components are removed from the initial conditions, while a new tensor of inertia is computed, in which only the forced components are kept, and the amplitude are corrected to consider the frequency-dependency of the Love number, performing this way a digital filtering.

\par The frequency analysis algorithm I use is based on NAFF (see \citep{l1993} for the method, and \citep{l2005} for the convergence proofs), with a refinement suggested by \citep{c1998} 
consisting in iterating the process to improve the accuracy of the determination. Correcting the initial condition to remove free oscillations has been used in different contexts these last years
\citep{lg2000,n2009,clcmu2010,rrc2011,d2011,hsl2014}, and the convergence of this algorithm is addressed in \citep{ndc2011}. \citet{ntt2015} propose a similar algorithm for the computation of periodic
orbits of artificial satellites of the Earth, but in using de Prony's frequency analysis \citep{d1795,tnt2014}. In our specific case, the algorithm goes further, in altering not only the initial conditions of
the solutions, but also the coefficients intervening in the equations.

\item A new numerical solution is computed, and the process is reiterated until convergence is reached. Then the observable quantities, such as the longitudinal librations and the obliquity,
are expressed as outputs.

\end{enumerate}

\subsection{Convergence}

\par For this algorithm to converge in the mathematical sense, it must at least have an attractive fixed point. This means that

\begin{itemize}
\item there should be a solution left unaltered by the algorithm, i.e. a set of time evolutions of the 6 elements of the tensor of inertia and of the 6 rotational quantities, which would both
the input and the output of the algorithm,
\item in starting from a solution, which would be close enough from this fixed point, i.e. in a neighbourhood, the algorithm should converge.
\end{itemize}

I here address the issue of the convergence in the planar eccentric problem, in the absence of planetary perturbations, and with a uniformly precessing pericentre. In that case, the only non-null elements of the tensor 
of inertia would be $I_{11}$, $I_{12}$, $I_{22}$ and $I_{33}$, and the rotational quantities would reduce to the spin angle $p$, and the norm of the normalized angular momentum $P=G/(nC)$. We would have a 
constant eccentricity $e$, a null inclination $I$, $\lambda(t)=nt$ and $\varpi(t)=\dot{\varpi}t$, $n$ and $\dot{\varpi}$ being constant. I set $n'=n-\dot{\varpi}$. The time origin can be conveniently chosen to 
give $\lambda(0)=\varpi(0)=0$.

\par I here use the Hamiltonian formulation. This provides us an opportunity to check the results obtained in Sect.~\ref{sec:longilibr} with the non-Hamiltonian formulation.

\subsubsection{In the elastic case}

\par We start from the following initial condition (from Eqs.~\ref{eq:I11freq}, \ref{eq:I12freq}, \ref{eq:I22freq} and \ref{eq:I33freq} for the tensor of inertia, and from the synchronous rotation for $p$ and $P$):

\begin{eqnarray}
I_{11}(t) & = & \frac{8\pi}{3}\rho X_4+\frac{n^2R^2}{g}\pi\rho X_3 \left(h_2(\chi_0)\left(-\frac{112}{45}+\frac{16}{5}e^2\right)-\frac{32}{5}h_2(\chi_1)e\cos(\lambda-\varpi)-16h_2(\chi_2)e^2\cos2(\lambda-\varpi)\right), \label{eq:I11conv} \\
I_{12}(t) & = & -\frac{4}{5}\frac{n^2R^2}{g}\pi\rho X_3\left(8h_2(\chi_1)e\sin(\lambda-\varpi)+17h_2(\chi_2)e^2\sin2(\lambda-\varpi)\right), \label{eq:I12conv} \\
I_{22}(t) & = & \frac{8\pi}{3}\rho X_4+\frac{n^2R^2}{g}\pi\rho X_3 \left(h_2(\chi_0)\left(\frac{32}{45}-\frac{24}{5}e^2\right)+\frac{16}{5}h_2(\chi_1)e\cos(\lambda-\varpi)+\frac{56}{5}h_2(\chi_2)e^2\cos2(\lambda-\varpi)\right), \label{eq:I22conv} \\
I_{33}(t) & = & \frac{8\pi}{3}\rho X_4+\frac{n^2R^2}{g}\pi\rho X_3 \left(h_2(\chi_0)\left(\frac{16}{9}+\frac{8}{5}e^2\right)+\frac{16}{5}h_2(\chi_1)e\cos(\lambda-\varpi)+\frac{24}{5}h_2(\chi_2)e^2\cos2(\lambda-\varpi)\right), \label{eq:I33conv} \\
p(t) & = & \lambda(t)-\pi, \label{eq:ppconv} \\
P & = & 1. \label{eq:GPconv}
\end{eqnarray}
This initial condition could be improved to consider the librations in $p$ and $P$ (Eqs.~\ref{eq:ppconv} \& \ref{eq:GPconv}). This inclusion would reduce the amplitude of the free oscillations, that the frequency analysis
would remove.

The Hamilton equations (\ref{eq:eqhamilton}) give

\begin{eqnarray}
\frac{dp}{dt} & = & nP, \label{eq:dppconv} \\
\frac{dP}{dt} & = & \frac{3}{2}n\left(2\gamma_4\sin(2f+2\varpi-2p)-\gamma_5\cos(2f+2\varpi-2p)\right), \label{eq:dGPconv}
\end{eqnarray}
$\gamma_2$ and $\gamma_3$ being null in the planar case, and $\gamma_1$ disappearing, the associated coefficient, i.e. $x^2+y^2$, being constant. I now set $p=\sigma+\lambda-\pi$, and from
$dP/dt=\ddot{p}/n=\ddot{\sigma}/n$, the libration angle $\sigma(t)$ is a solution of the following ODE:

\begin{equation}
\label{eq:ODEconv}
\ddot{\sigma}+\omega_0^2\sigma=\alpha_1\sin(\lambda-\varpi)+\alpha_2\sin2(\lambda-\varpi),
\end{equation}
with

\begin{eqnarray}
\omega_0^2 & = & \frac{48}{5}\frac{n^4R^2}{g\mathcal{C}}\pi\rho X_3\left(h_2(\chi_0)(1-5e^2)+2e^2h_2(\chi_1)\right), \label{eq:omega0conv} \\
\alpha_1 & = & \frac{96}{5}e\frac{n^4R^2}{g\mathcal{C}}\pi\rho X_3\left(h_2(\chi_0)-h_2(\chi_1)\right) \label{eq:alpha1conv} \\
& \approx & 2e\omega_0^2\frac{h_2(\chi_0)-h_2(\chi_1)}{h_2(\chi_0)}, \label{eq:alpha1conv2} \\
\alpha_2 & = & \frac{204}{5}e^2\frac{n^4R^2}{g\mathcal{C}}\pi\rho X_3\left(h_2(\chi_0)-h_2(\chi_2)\right) \label{eq:alpha2conv} \\
& \approx & \frac{17}{4}e^2\omega_0^2\frac{h_2(\chi_0)-h_2(\chi_2)}{h_2(\chi_0)}. \label{eq:alpha2conv2}
\end{eqnarray}
This ODE is straightforwardly solved as

\begin{eqnarray}
p(t) & = & \lambda-\pi+\frac{\alpha_1}{\omega_0^2-n'^2}\sin(\lambda-\varpi)+\frac{\alpha_2}{\omega_0^2-4n'^2}\sin2(\lambda-\varpi)+A\cos(\omega_0t+\Psi) \label{eq:ppit1conv} \\
P(t) & = & 1+\frac{\alpha_1n'}{n\left(\omega_0^2-n'^2\right)}\cos(\lambda-\varpi)+\frac{2\alpha_2n'}{n\left(\omega_0^2-4n'^2\right)}\cos2(\lambda-\varpi)-A\frac{\omega_0}{n}\sin(\omega_0t+\Psi), \label{eq:GPit1conv}
\end{eqnarray}
$A$ and $\Psi$ being the amplitude and phase of a free oscillation, which reflects the departure of the chosen initial condition for $p$ and $P$ from the exact equilibrium. In particular, the
third degree in the eccentricity has been neglected.

The resulting tensor of inertia is, after digital filtering:

\begin{eqnarray}
I_{11}(t) & = & \frac{8\pi}{3}\rho X_4+\frac{n^2R^2}{g}\pi\rho X_3 \left(h_2(\chi_0)\left(-\frac{112}{45}+\frac{16}{5}e^2\right)-\frac{32}{5}h_2(\chi_1)e\cos(\lambda-\varpi)-16h_2(\chi_2)e^2\cos2(\lambda-\varpi)\right) \nonumber \\
& & -\frac{32}{45}e\frac{n'}{n}h_2(\chi_1)\left(h_2(\chi_1)-h_2(\chi_0)\right)\frac{\omega_0^2}{n'^2-\omega_0^2}\frac{n^2R^2\pi\rho X_3}{gh_2(\chi_0)}\cos(\lambda-\varpi), \label{eq:I11bconv} \\
I_{12}(t) & = & -\frac{4}{5}\frac{n^2R^2}{g}\pi\rho X_3\left(8h_2(\chi_1)e\sin(\lambda-\varpi)+17h_2(\chi_2)e^2\sin2(\lambda-\varpi)\right) \nonumber \\
& & -\frac{32}{5}eh_2(\chi_1)\left(h_2(\chi_0)-h_2(\chi_1)\right)\frac{n^2\omega_0^2}{n'^2-\omega_0^2}\frac{R^2\pi\rho X_3}{gh_2(\chi_0)}\sin(\lambda-\varpi), \label{eq:I12bconv} \\
I_{22}(t) & = & \frac{8\pi}{3}\rho X_4+\frac{n^2R^2}{g}\pi\rho X_3 \left(h_2(\chi_0)\left(\frac{32}{45}-\frac{24}{5}e^2\right)+\frac{16}{5}h_2(\chi_1)e\cos(\lambda-\varpi)+\frac{56}{5}h_2(\chi_2)e^2\cos2(\lambda-\varpi)\right) \nonumber \\
& & -\frac{32}{45}e\frac{n'}{n}h_2(\chi_1)\left(h_2(\chi_1)-h_2(\chi_0)\right)\frac{\omega_0^2}{n'^2-\omega_0^2}\frac{n^2R^2\pi\rho X_3}{gh_2(\chi_0)}\cos(\lambda-\varpi), \label{eq:I22bconv} \\
I_{33}(t) & = & \frac{8\pi}{3}\rho X_4+\frac{n^2R^2}{g}\pi\rho X_3 \left(h_2(\chi_0)\left(\frac{16}{9}+\frac{8}{5}e^2\right)+\frac{16}{5}h_2(\chi_1)e\cos(\lambda-\varpi)+\frac{24}{5}h_2(\chi_2)e^2\cos2(\lambda-\varpi)\right) \nonumber \\
& & -\frac{1024}{75}eh_2(\chi_1)\left(h_2(\chi_0)-h_2(\chi_1)\right)\frac{n^5n'}{n'^2-\omega_0^2}\frac{R^4\pi^2\rho^2X_3^2}{g\mathcal{C}}\cos(\lambda-\varpi), \label{eq:I33bconv}
\end{eqnarray}
which itself gives this new ODE ruling $\sigma$:

\begin{equation}
\label{eq:ODEbconv}
\ddot{\sigma}+\omega_0^2\sigma = \left(\alpha_1-\alpha'_1\right)\sin(\lambda-\varpi)+\left(\alpha_2-\alpha'_2\right)\sin2(\lambda-\varpi),
\end{equation}
with 

\begin{eqnarray}
\alpha'_1 & = & 2e\frac{h_2(\chi_1)\left(h_2(\chi_1)-h_2(\chi_0)\right)}{h_2(\chi_0)^2}\frac{\omega_0^4}{n'^2-\omega_0^2}, \\ \label{eq:alpha1bconv}
\alpha'_2 & = & 3e^2\frac{h_2(\chi_2)\left(h_2(\chi_1)-h_2(\chi_0)\right)}{h_2(\chi_0)^2}\frac{\omega_0^4}{n'^2-\omega_0^2}. \label{eq:alpha2bconv}
\end{eqnarray}

In these expressions, the amplitudes having been expanded up to the second degree in the eccentricity, and the first degree in $1/\mathcal{C}$. Moreover, the squared amplitude of the free oscilation $A^2$ affects the constant term. 
This quantity can be neglected, since $A$ is originally small, and the other iterations will correct the initial conditions to make it disappear.

\par We can see that our original solution for the tensor of inertia and the rotational quantities (Eq.~\ref{eq:I11conv} to \ref{eq:GPconv}) is not exactly a fixed point. It could anyway be an acceptable one
if the algorithm does not diverge too much. This could be checked in evaluating the ratios $\left|\alpha'_1/\alpha_1\right|$ and $\left|\alpha'_2/\alpha_2\right|$.

\par We have

\begin{eqnarray}
\left|\frac{\alpha'_1}{\alpha_1}\right| & \approx & \frac{h_2(\chi_1)}{h_2(\chi_0)}\frac{\omega_0^2}{|n'^2-\omega_0^2|}, \\ \label{eq:rapalfa1}
\left|\frac{\alpha'_2}{\alpha_2}\right| & \approx & \frac{17}{12}\frac{h_2(\chi_2)}{h_2(\chi_0)}\frac{\omega_0^2}{|n'^2-\omega_0^2|}. \label{eq:rapalfa2}
\end{eqnarray}

\par We can see that, as in many cases in celestial mechanics, a resonance is a potential problem, i.e. if $n'$ and $\omega_0$ are close. This non-resonant condition can be more 
difficult to fulfill in many dimensions. Usually $\omega_0$ is small with respect to $n'$, which infers that the two ratios $\left|\alpha'_1/\alpha_1\right|$ and $\left|\alpha'_2/\alpha_2\right|$
are small.

\par For the parameters relevant to the crust of Europa, these ratios are of the order of $10^{-3}$, which gives the order of magnitude of the uncertainty on the amplitude given by this algorithm, 
for a limited number of iterations. The results presented in Sect.~\ref{sec:results} have been obtained after 3 iterations.

\subsubsection{Introduction of the dissipation}

\par The initial condition is now 

\begin{eqnarray}
I_{11}(t) & = & \frac{8\pi}{3}\rho X_4+\frac{n^2R^2}{g}\pi\rho X_3 \left(h_2(\chi_0)\left(-\frac{112}{45}+\frac{16}{5}e^2\right)-\frac{32}{5}h_2(\chi_1)e\cos(\lambda-\varpi)-\frac{32}{5}\left(\frac{h_2}{Q_h}\right)(\chi_1)e\sin(\lambda-\varpi)\right. \nonumber \\
& & \left.-16h_2(\chi_2)e^2\cos2(\lambda-\varpi)-16\left(\frac{h_2}{Q_h}\right)(\chi_2)e^2\sin2(\lambda-\varpi)\right), \label{eq:I11diss} \\
I_{12}(t) & = & -\frac{4}{5}\frac{n^2R^2}{g}\pi\rho X_3\left(8h_2(\chi_1)e\sin(\lambda-\varpi)-8\left(\frac{h_2}{Q_h}\right)(\chi_1)e\cos(\lambda-\varpi)\right. \nonumber \\
& & \left.+17h_2(\chi_2)e^2\sin2(\lambda-\varpi)-17\left(\frac{h_2}{Q_h}\right)(\chi_2)e^2\cos2(\lambda-\varpi)\right), \label{eq:I12diss} \\
I_{22}(t) & = & \frac{8\pi}{3}\rho X_4+\frac{n^2R^2}{g}\pi\rho X_3 \left(h_2(\chi_0)\left(\frac{32}{45}-\frac{24}{5}e^2\right)+\frac{16}{5}h_2(\chi_1)e\cos(\lambda-\varpi)+\frac{16}{5}\left(\frac{h_2}{Q_h}\right)(\chi_1)e\sin(\lambda-\varpi)\right. \nonumber \\
& & \left.+\frac{56}{5}h_2(\chi_2)e^2\cos2(\lambda-\varpi)+\frac{56}{5}\left(\frac{h_2}{Q_h}\right)(\chi_2)e^2\sin2(\lambda-\varpi)\right), \label{eq:I22diss} \\
I_{33}(t) & = & \frac{8\pi}{3}\rho X_4+\frac{n^2R^2}{g}\pi\rho X_3 \left(h_2(\chi_0)\left(\frac{16}{9}+\frac{8}{5}e^2\right)+\frac{16}{5}h_2(\chi_1)e\cos(\lambda-\varpi)+\frac{16}{5}\left(\frac{h_2}{Q_h}\right)(\chi_1)e\sin(\lambda-\varpi)\right. \nonumber \\
& & \left.+\frac{24}{5}h_2(\chi_2)e^2\cos2(\lambda-\varpi)+\frac{24}{5}\left(\frac{h_2}{Q_h}\right)(\chi_2)e^2\sin2(\lambda-\varpi)\right), \label{eq:I33diss} \\
p(t) & = & \lambda(t)-\pi, \label{eq:ppdiss} \\
P & = & 1, \label{eq:GPdiss}
\end{eqnarray}
which gives the following ODE:

\begin{equation}
\label{eq:ODEdiss}
\ddot{\sigma}+\omega_0^2\sigma=\beta_0+\alpha_1\sin(\lambda-\varpi)+\beta_1\cos(\lambda-\varpi)+\alpha_2\sin2(\lambda-\varpi)+\beta_2\cos2(\lambda-\varpi),
\end{equation}
with

\begin{eqnarray}
\beta_0 & = & \frac{288}{5}\frac{n^4R^2}{g\mathcal{C}}\pi\rho X_3 \left(\frac{h_2}{Q_h}\right)(\chi_1) \label{eq:beta0diss} \\
& \approx & 6\omega_0^2\frac{(h_2/Q_h)(\chi_1)}{h_2(\chi_0)}, \label{eq:beta0diss2} \\
\beta_1 & = & \frac{96}{5}\frac{n^4R^2}{g\mathcal{C}}\pi\rho X_3 \left(\frac{h_2}{Q_h}\right)(\chi_1) \label{eq:beta1diss} \\
& \approx & 2\omega_0^2\frac{(h_2/Q_h)(\chi_1)}{h_2(\chi_0)}, \label{eq:beta1diss2} \\
\beta_2 & = & \frac{204}{5}\frac{n^4R^2}{g\mathcal{C}}\pi\rho X_3 \left(\frac{h_2}{Q_h}\right)(\chi_2) \label{eq:beta2diss} \\
& \approx & \frac{17}{4}\omega_0^2\frac{(h_2/Q_h)(\chi_2)}{h_2(\chi_0)}, \label{eq:beta2diss2}
\end{eqnarray}
$\omega_0$, $\alpha_1$ and $\alpha_2$ being unchanged (Eqs.~\ref{eq:omega0conv} to \ref{eq:alpha2conv}). The solutions of this ODE are:

\begin{eqnarray}
p(t) & = & \lambda-\pi+\frac{\beta_0^2}{\omega_0^2}+\frac{\sqrt{\alpha_1^2+\beta_1^2}}{\omega_0^2-n'^2}\sin(\lambda-\varpi+\phi_1)+\frac{\sqrt{\alpha_2^2+\beta_2^2}}{\omega_0^2-4n'^2}\sin2(\lambda-\varpi+\phi_2)+A\cos(\omega_0t+\Psi) \label{eq:ppit1diss} \\
P(t) & = & 1+\frac{n'\sqrt{\alpha_1^2+\beta_1^2}}{n\left(\omega_0^2-n'^2\right)}\cos(\lambda-\varpi+\phi_1)+\frac{2n'\sqrt{\alpha_2^2+\beta_2^2}}{n\left(\omega_0^2-4n'^2\right)}\cos2(\lambda-\varpi)-A\frac{\omega_0}{n}\sin(\omega_0t+\Psi), \label{eq:GPit1diss}
\end{eqnarray}
with

\begin{eqnarray}
\tan\phi_1 & = & \frac{(h_2/Q_h)(\chi_1))}{h_2(\chi_1)-h_2(\chi_0)},  \label{eq:tphi1diss} \\
\tan\phi_2 & = & \frac{(h_2/Q_h)(\chi_2))}{h_2(\chi_2)-h_2(\chi_0)}.  \label{eq:tphi2diss}
\end{eqnarray}
A constant term appears in $p(t)$, which is equal to $\sigma_0=\beta_0^2/\omega_0^2=6e^2\frac{(h_2/Q_h)(\chi_1)}{h_2(\chi_0)}$, consistently with Eq.~\ref{eq:sigma0}. These phase angles 
are the critical outcome of the introduction of the dissipation, while the amplitudes are only marginally affected, being multiplied by the ratios 
$\frac{\sqrt{(h_2(\chi_i)-h_2(\chi_0))^2}+(h_2/Q_h)(\chi_i)^2}{h_2(\chi_i)-h_2(\chi_0)}$, which are very close to 1.

\par The solution (\ref{eq:ppit1diss},\ref{eq:GPit1diss}) gives the following tensor of inertia, after digital filtering and addition of the orthogonal, dissipative contribution:

\begin{eqnarray}
I_{11}(t) & = & \frac{8\pi}{3}\rho X_4+\frac{n^2R^2}{g}\pi\rho X_3 \left(h_2(\chi_0)\left(-\frac{112}{45}+\frac{16}{5}e^2+\frac{64}{5}e^2\left(h_2(\chi_1)-h_2(\chi_0)\right)\right)\right. \nonumber \\
& & \left.-\frac{32}{5}e\left(h_2(\chi_1)\cos(\lambda-\varpi)+\left(\frac{h_2}{Q_h}\right)(\chi_1)\sin(\lambda-\varpi)\right)-16e^2\left(h_2(\chi_2)\cos2(\lambda-\varpi)+\left(\frac{h_2}{Q_h}\right)(\chi_2)\sin2(\lambda-\varpi)\right)\right) \label{eq:I11bdiss} \\
& & +\frac{32}{45}e\frac{n'}{n}\frac{\omega_0^2}{n'^2-\omega_0^2}\frac{n^2R^2}{gh_2(\chi_0)}\pi\rho X_3\sqrt{(h_2(\chi_0)-h_2(\chi_1))^2+(h_2/Q_h)(\chi_1)^2} \nonumber \\
& & \times\left(h_2(\chi_1)\cos(\lambda-\varpi+\phi_1)+\left(\frac{h_2}{Q_h}\right)(\chi_1)\sin(\lambda-\varpi+\phi_1)\right), \nonumber \\
I_{12}(t) & = & \frac{48}{5}\frac{n^2R^2}{g}\pi\rho X_3e^2\left(\frac{h_2}{Q_h}\right)(\chi_1)\left(2-\frac{\omega_0^2}{n'^2-\omega_0^2}\right)-\frac{4}{5}\frac{n^2R^2}{g}\pi\rho X_3\left(8h_2(\chi_1)e\sin(\lambda-\varpi)-8\left(\frac{h_2}{Q_h}\right)(\chi_1)e\cos(\lambda-\varpi)\right. \nonumber \\
& & \left.+17h_2(\chi_2)e^2\sin2(\lambda-\varpi)-17\left(\frac{h_2}{Q_h}\right)(\chi_2)e^2\cos2(\lambda-\varpi)\right) \label{eq:I12bdiss} \\
& & -\frac{32}{5}\frac{\omega_0^2}{n'^2-\omega_0^2}e\frac{n^2R^2}{gh_2(\chi_0)}\pi\rho X_3\sqrt{(h_2(\chi_0)-h_2(\chi_1))^2+(h_2/Q_h)(\chi_1)^2} \nonumber \\
& & \times\left(h_2(\chi_1)\sin(\lambda-\varpi+\phi_1)-\left(\frac{h_2}{Q_h}\right)(\chi_1)\cos(\lambda-\varpi+\phi_1)\right), \nonumber \\
I_{22}(t) & = & \frac{8\pi}{3}\rho X_4+\frac{n^2R^2}{g}\pi\rho X_3 \left(h_2(\chi_0)\left(\frac{32}{45}-\frac{24}{5}e^2-\frac{64}{5}e^2\left(h_2(\chi_1)-h_2(\chi_0)\right)\right)\right. \nonumber \\
& & \left.+\frac{16}{5}e\left(h_2(\chi_1)\cos(\lambda-\varpi)+\left(\frac{h_2}{Q_h}\right)(\chi_1)\sin(\lambda-\varpi)\right)+\frac{56}{5}e^2\left(h_2(\chi_2)\cos2(\lambda-\varpi)+\left(\frac{h_2}{Q_h}\right)(\chi_2)\sin2(\lambda-\varpi)\right)\right) \label{eq:I22bdiss} \\
& & +\frac{32}{45}e\frac{n'}{n}\frac{\omega_0^2}{n'^2-\omega_0^2}\frac{n^2R^2}{gh_2(\chi_0)}\pi\rho X_3\sqrt{(h_2(\chi_0)-h_2(\chi_1))^2+(h_2/Q_h)(\chi_1)^2} \nonumber \\
& & \times\left(h_2(\chi_1)\cos(\lambda-\varpi+\phi_1)+\left(\frac{h_2}{Q_h}\right)(\chi_1)\sin(\lambda-\varpi+\phi_1)\right), \nonumber \\
I_{33}(t) & = & \frac{8\pi}{3}\rho X_4+\frac{n^2R^2}{g}\pi\rho X_3 \left(h_2(\chi_0)\left(\frac{16}{9}+\frac{8}{5}e^2\right)+\frac{16}{5}e\left(h_2(\chi_1)\cos(\lambda-\varpi)+\left(\frac{h_2}{Q_h}\right)(\chi_1)\sin(\lambda-\varpi)\right)\right. \nonumber \\
& & \left.+\frac{24}{5}e^2\left(h_2(\chi_2)\cos2(\lambda-\varpi)+\left(\frac{h_2}{Q_h}\right)(\chi_2)\sin2(\lambda-\varpi)\right)\right) \label{eq:I33bdiss} \\
& & -\frac{64}{45}e\frac{n'}{n}\frac{\omega_0^2}{n'^2-\omega_0^2}\frac{n^2R^2}{gh_2(\chi_0)}\pi\rho X_3\sqrt{(h_2(\chi_0)-h_2(\chi_1))^2+(h_2/Q_h)(\chi_1)^2} \nonumber \\
& & \times\left(h_2(\chi_1)\cos(\lambda-\varpi+\phi_1)+\left(\frac{h_2}{Q_h}\right)(\chi_1)\sin(\lambda-\varpi+\phi_1)\right), \nonumber 
\end{eqnarray}
which gives the following ODE:

\begin{equation}
\label{eq:ODEbdiss}
\ddot{\sigma}+\omega_0^2\sigma = \beta'_0+\alpha_1\sin(\lambda-\varpi)+\beta_1\cos(\lambda-\varpi)+\zeta_1\sin(\lambda-\varpi+\phi_1)+\kappa_1\cos(\lambda-\varpi+\phi_1),
\end{equation}
the second harmonic in $\lambda-\varpi$ having been removed, since it is not critical in the process of convergence. This is why $\alpha_2$ and $\beta_2$ do not appear anymore.
The frequency of the proper oscillations $\omega_0$ is only marginally altered, this is why I do not introduce another notation here. However, the phases for the harmonic
$0$, previously called $\sigma_0$, and now called $\sigma'_0=\beta'_0/\omega_0^2$, and of the harmonic $1$ $\phi_1$, are significantly altered.

We have

\begin{equation}
\label{eq:betap0}
\beta'_0 = \frac{144}{5}\frac{n^4}{n'^2-\omega_0^2}\frac{R^2\pi\rho X_3}{gC}e^2\left(\left(\frac{h_2}{Q_h}\right)(\chi_1)\left(4n'^2-5\omega_0^2+\omega_0^2\cos\phi_1\right)-h_2(\chi_1)\omega_0^2\sin\phi_1\right),
\end{equation}
which can be approximated to $2\sigma_0$ if we neglect $\omega_0^2$ with respect to $n^2$ and $n'^2$, and assume $n\approx n'$. This means that the phase shift $\sigma_0$, which is considered 
in the solution of $p$ coming from the first iteration of the algorithm, is supplemented by itself. Physically, this is the phase lag introduced by the dissipation, and it seems to make sense
to artificially compensate this lag at each iteration. Unfortunately, the problem of phase shift holds for each harmonic. The new phase $\phi_1$ will be ruled not only by $\alpha_1$ and
$\beta_1$, but also by the coefficients multiplying $\sin(\lambda-\varpi+\phi_1)$ and $\cos(\lambda-\varpi+\phi_1)$, i.e.

\begin{eqnarray}
\zeta_1 & = & -\frac{96}{5}e\frac{\omega^2n^4}{n'^2-\omega_0^2}\frac{R^2\pi\rho X_3}{gC}\frac{h_2(\chi_1)}{h_2(\chi_0)}, \label{eq:zeta1} \\
\kappa_1 & = & -\frac{96}{5}e\frac{\omega^2n^4}{n'^2-\omega_0^2}\frac{R^2\pi\rho X_3}{gC}\frac{(h_2/Q_h)(\chi_1)}{h_2(\chi_0)}. \label{eq:kappa_1}
\end{eqnarray}

So, the required correction is a different one. Such a problem, due to delayed quantities, has already been addressed in the context of fitting ephemerides of the Moon to Lunar Laser Ranging \citep[e.g.]{m2011,b2016}. In particular \citet{m2011} tested different ways to minimize the global error, but this did not permit to simulate the different phase lags accurately.

\par It can be considered that the algorithm converges \emph{enough} in the absence of dissipation, in the sense that the resulting amplitudes will be reliable. However, this cannot be used for a dissipative dynamics, since the information is essentially contained in the phase lags. In the following, I will use numerical simulations only in the elastic case.

\section{Results}\label{sec:results}

\par I here present outputs of simulation of the rotation of an elastic crust, adapted to Europa. These results have been obtained in simulating the rotation over 400 years with a timestep of 0.01
orbital period of Europa, i.e. $\approx3.551$ d. The frequency analyses have been made with 262,144 points, evenly spaced, separated by $\approx0.557$ day. The total duration of the integration
allows to distinguish constant contributions from sinusoids with a period up to 200 years, and the limited timestep in the frequency analysis permits to detect periods as small as twice the timestep,
i.e. $\approx1.2$ d. I present only simulations obtained with the Hamiltonian code, but the non-Hamiltonian one is in perfect agreement.

\subsection{The resulting tensor of inertia}

\par The Tab.~\ref{tab:I11num} to \ref{tab:I33num} give the resulting tensor of inertia for a typical crust, i.e. $\rho = 800$ kg.m$^{-3}$, $\eth = 14$ km, 
$h_2(\chi_0)=2.03793$, $k_2(\chi_0) = 1.03842$, $\tau_M=\tau_A=3\times10^5$ s, and $\alpha = 0.3$. For every amplitude, I compare the analytical formula (Eqs.~\ref{eq:I11freq} to \ref{eq:I33freq}))
with the one given by the numerical simulation.

\begin{table}
\centering
\caption{Frequency decomposition of $I_{11}$. The unit is kg.m$^2$.\label{tab:I11num}}
\begin{tabular}{cccccccc}
Analytical & Numerical & $\lambda_2$ & $\ascnode_2$ & $\varpi_2$ &  $\varpi_3$ & $\lambda_1$ & Period \\
\hline
$5.46411\times10^{26}$ & $5.46411\times10^{26}$ & -  & -  & -  & -  & - & $\infty$ \\
$6.89524\times10^{21}$ & $6.88936\times10^{20}$ & -1 & -  & -  & -  & 1 & $3.5255$ d \\
-                      & $1.57479\times10^{20}$ & 1  & -  & -  & -1 & - & $3.5514$ d \\
$1.60377\times10^{20}$ & $1.58310\times10^{19}$ & -2 & -  & -  & -  & 2 & $1.7627$ d \\
-                      & $1.45809\times10^{19}$ & 1  & -  & -1 & -  & - & $3.5525$ d \\
$9.74188\times10^{17}$ & -                      & 2  & -2 & -  & -  & - & $1.7750$ d \\
\hline
\end{tabular}
\end{table}

\begin{table}
\centering
\caption{Frequency decomposition of $I_{12}$.\label{tab:I12num}}
\begin{tabular}{cccccccc}
Analytical & Numerical & $\lambda_2$ & $\ascnode_2$ & $\varpi_2$ &  $\varpi_3$ &  $\lambda_1$ & Period \\
\hline
$6.89524\times10^{21}$ & $6.82478\times10^{21}$ & -1 & -  & -  & -  & 1 & $3.5255$ d \\
-                      & $1.57853\times10^{20}$ &  1 & -  & -  & -1 & - & $3.5514$ d \\
-                      & $1.46158\times10^{20}$ &  1 & -  & -1 & -  & - & $3.5525$ d \\
$1.36321\times10^{20}$ & $1.34667\times10^{20}$ & -2 & -  & -  & -  & 2 & $1.7627$ d \\
$4.87094\times10^{17}$ & -                      &  2 & -2 & -  & -  & - & $1.7750$ d \\
\hline
\end{tabular}
\end{table}

\begin{table}
\centering
\caption{Frequency decomposition of $I_{13}$.\label{tab:I13num}}
\begin{tabular}{cccccccc}
Analytical & Numerical & $\lambda_2$ & $\nu$ & $\ascnode_2$ & $\ascnode_3$ & $\lambda_1$ & Period \\
\hline
$8.49742\times10^{20}$ & $8.50348\times10^{20}$ & 1  & -  & -1 & -  & - & $3.5500$ d \\
$1.97634\times10^{19}$ & $1.96972\times10^{19}$ & -  & -  & -1 & -  & 1 & $1.7689$ d \\
$6.47876\times10^{18}$ & $7.28366\times10^{18}$ & -  & -1 & -1 & -  & - & $1.3943$ y \\
-                      & $4.79594\times10^{18}$ & 1  & -  & -  & -1 & - & $3.5509$ d \\
\hline
\end{tabular}
\end{table}

\begin{table}
\centering
\caption{Frequency decomposition of $I_{22}$.\label{tab:I22num}}
\begin{tabular}{ccccccc}
Analytical & Numerical & $\lambda_2$ & $\varpi_2$ &  $\varpi_3$ & $\lambda_1$ & Period \\
\hline
$5.47083\times10^{26}$ & $5.47083\times10^{26}$ & -  & -  & -  & - & $\infty$ \\
$3.44762\times10^{21}$ & $3.44764\times10^{21}$ & -1 & -  & -  & 1 & $3.5255$ d \\
$1.12264\times10^{20}$ & $1.10627\times10^{20}$ & -2 & -  & -  & 2 & $1.7627$ d \\
-                      & $7.87896\times10^{19}$ & 1  & -  & -1 & - & $3.5514$ d \\
-                      & $7.29472\times10^{19}$ & 1  & -1 & -  & - & $3.5525$ d \\
\hline
\end{tabular}
\end{table}

\begin{table}
\centering
\caption{Frequency decomposition of $I_{23}$.\label{tab:I23num}}
\begin{tabular}{ccccccc}
Analytical & Numerical & $\lambda_2$ & $\nu$ & $\ascnode_2$ & $\lambda_1$ & Period \\
\hline
$1.29575\times10^{19}$ & $1.45558\times10^{19}$ & -  & -1 & -1 & - & $1.3943$ y \\
$7.90538\times10^{18}$ & $7.84040\times10^{18}$ & -  & -  & -1 & 1 & $1.7689$ d \\
\hline
\end{tabular}
\end{table}

\begin{table}
\centering
\caption{Frequency decomposition of $I_{33}$.\label{tab:I33num}}
\begin{tabular}{ccccccccc}
Analytical & Numerical & $\lambda_2$ & $\ascnode_2$ & $\varpi_2$ &  $\varpi_3$ & $\ascnode_3$ & $\lambda_1$ & Period \\
\hline
$5.47307\times10^{26}$ & $5.47307\times10^{26}$ & -  & -  & -  & -  & - & - & $\infty$ \\
$3.44762\times10^{21}$ & $3.44172\times10^{21}$ & -1 & -  & -  & -  & - & 1 & $3.5255$ d \\
-                      & $7.86874\times10^{19}$ & 1  & -  & -  & -1 & - & - & $3.5514$ d \\
-                      & $7.27296\times10^{19}$ & 1  & -  & -1 & -  & - & - & $3.5525$ d \\
$4.81132\times10^{19}$ & $4.76856\times10^{19}$ & -2 & -  & -  & -  & - & 2 & $1.7627$ d \\
$9.74188\times10^{17}$ & -                      &  2 & -2 & -  & -  & - & - & $1.7750$ d \\
\hline
\end{tabular}
\end{table}

\par Not only the resulting amplitudes are in very good agreement with the theory, but other contributions appear, which are due to the gravitational perturbations
of the other Galilean satellites on the orbit of Europa. These contributions can have larger amplitudes that some given by the theory. Even it is possible to 
estimate them analytically, the use of numerical simulations with real orbital ephemerides permits to avoid many approximations required by an analytical derivation.
In particular, it prevents from forgetting some contributions, and it could even emphasize some resonant phenomena, even if there are none in this specific case.
This by far compensates the loss of accuracy due to the lack of mathematical convergence of the algorithm.

\subsection{Librations and obliquity}

\par I identified the longitudinal librations as the oscillations of the signal $p(t)$ after removal of a constant slope, and the latitudinal ones as the variations of the figure axis 
$\hat{f_3}$. The Tab.~\ref{tab:longi} \& \ref{tab:lati} give the main terms and the amplitudes for the typical case investigated in the previous section.

\begin{table}
\centering
\caption{Longitudinal librations. These quantities have been obtained after the numerical simulation of the rotation of a typical crust for Europa.\label{tab:longi}}
\begin{tabular}{c|ccccccccccc}
N &  Amplitude (arcmin)& $\lambda_2$ & $\nu$ & $\varpi_2$ & $\varpi_3$ & $\varpi_4$ & $\varpi_1$ & $\lambda_{\sun}$ & $\lambda_1$ & $\psi$ & Period \\
\hline
1 & $2.9947$           & -           &   1   & -          &   1        & -          & -          & -                & -           & -      & $1.3198$ y \\
2 & $1.5882$           & -           &   1   & -          & -          & 1          & -          & -                & -           & -      & $1.3295$ y \\
3 & $1.0743$           & -           &   1   & 1          & -          & -          & -          & -                & -           & -      & $1.2663$ y \\
4 & $0.6826$           & -           &   -   & -          & -          & -          & -          & -                & -           & 1      & $5.6387$ y \\
5 & $0.6311$           & -           &   -   & -          & -          & -          & -          & 1                & -           & -      & $11.863$ y \\
6 & $0.2222$           & -           &   1   & -          & -          & -          & 1          & -                & -           & -      & $1.1048$ y \\
\hline
\end{tabular}
\end{table}

\begin{table}
\centering
\caption{Latitudinal librations. They result from the same computation as the longitudinal ones (Tab.~\ref{tab:longi}). The amplitude of the first contribution 
is the obliquity.\label{tab:lati}}
\begin{tabular}{c|cccccc}
N &  Amplitude (arcmin) & $\lambda_2$ & $\ascnode_2$ & $\ascnode_3$ & $\lambda_1$ & Period \\
\hline
1 & $7.9432$            & 1           & -1           & -            & -           & $3.5500$ d \\
2 & $0.0736$            & -           & 1            & -            & -1          & $1.7689$ d \\
3 & $0.0637$            & 1           & -            & -1           & -           & $3.5509$ d \\
\hline
\end{tabular}
\end{table}

\par Once more, we can see that the numerical algorithm goes further than the analytical derivation, in permitting the identification of more terms, which are due to 
the planetary perturbations. The only exception is the amplitude of the diurnal librations, at the period $3.5255$ days, which are expected from Eq.~\ref{eq:sigma1} to have an amplitude half the 
one of the sixth term, i.e. 0.108 arcmin, hence too small to be numerically detected. A spectral analysis centered on this period shows a peak with an amplitude of 0.124 arcmin, but with a doubtful accuracy, since
other periodic contributions noise the analysis. I also computed these numbers for many models of crust of Europa. I considered 785 models, 
extracted from the set built in Sect.~\ref{sec:interior}.
The 9 periodic contributions identified in Tab.~\ref{tab:longi} \& \ref{tab:lati} have also been identified in the 785 simulations.

\begin{figure}
\centering
\begin{tabular}{cc}
\includegraphics[width=0.45\textwidth]{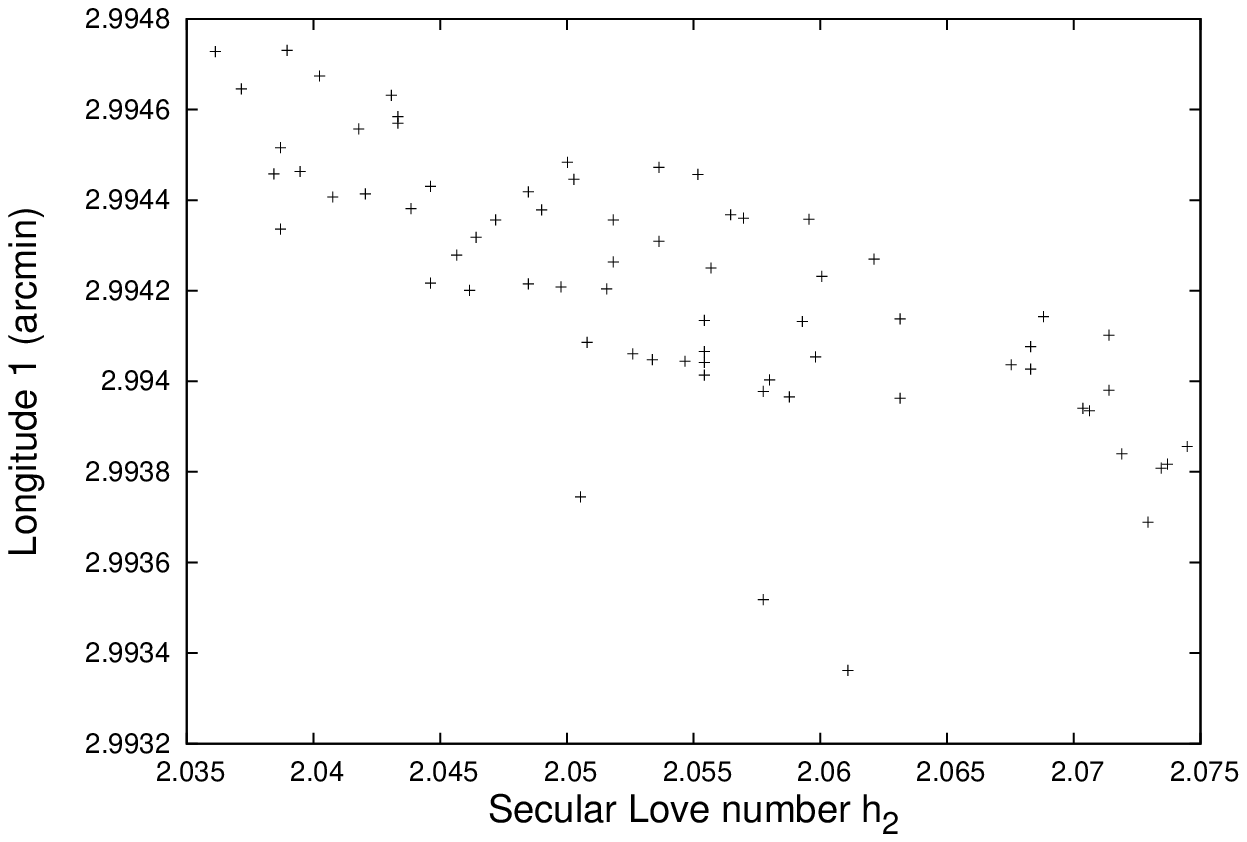} & \includegraphics[width=0.45\textwidth]{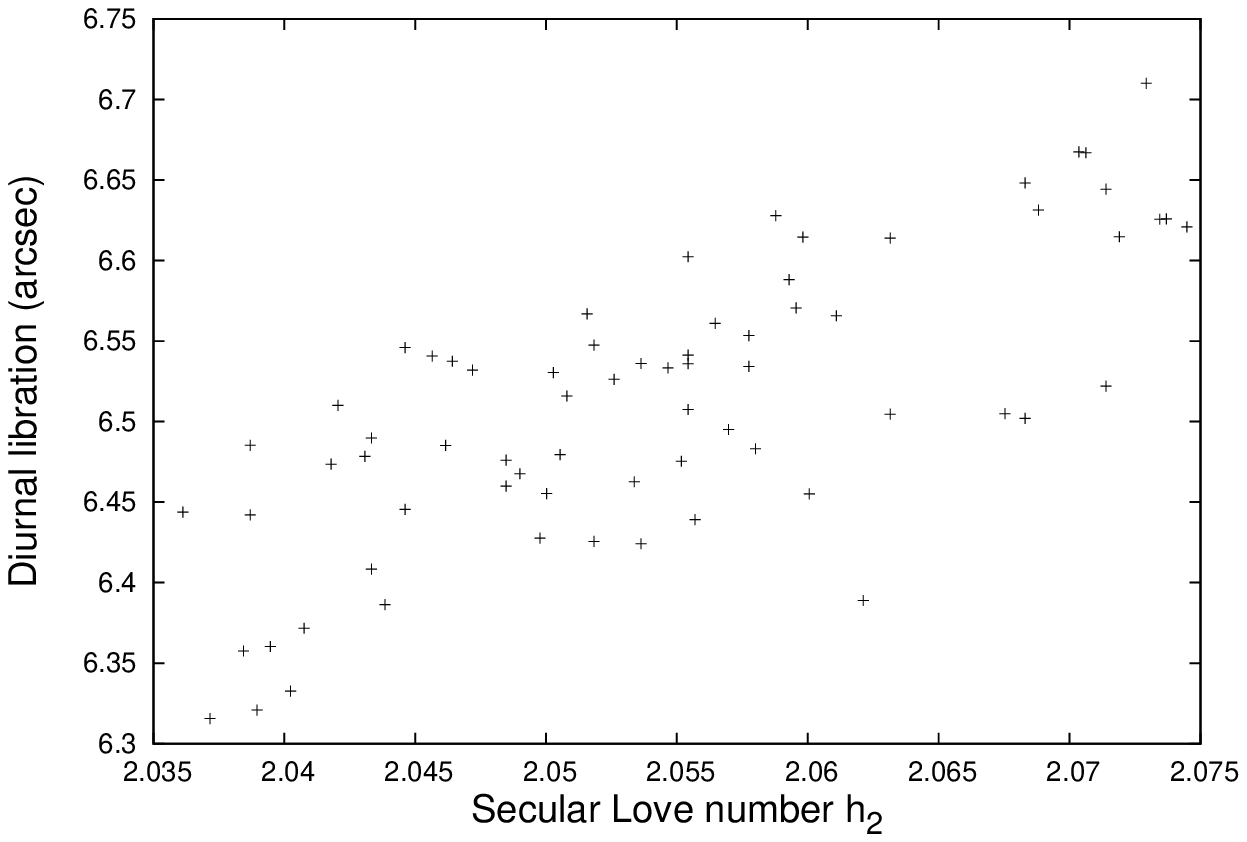}
\end{tabular}
\caption{Amplitude of the main term of the longitudinal librations of the crust of Europa, with period $1.3198$ year (left), and of the diurnal libration, i.e. at the period $3.52546$ days,
obtained with the Eq.~\ref{eq:sigma1} (right).\label{fig:longi}}
\end{figure}

\begin{figure}
\centering
\begin{tabular}{cc}
\includegraphics[width=0.45\textwidth]{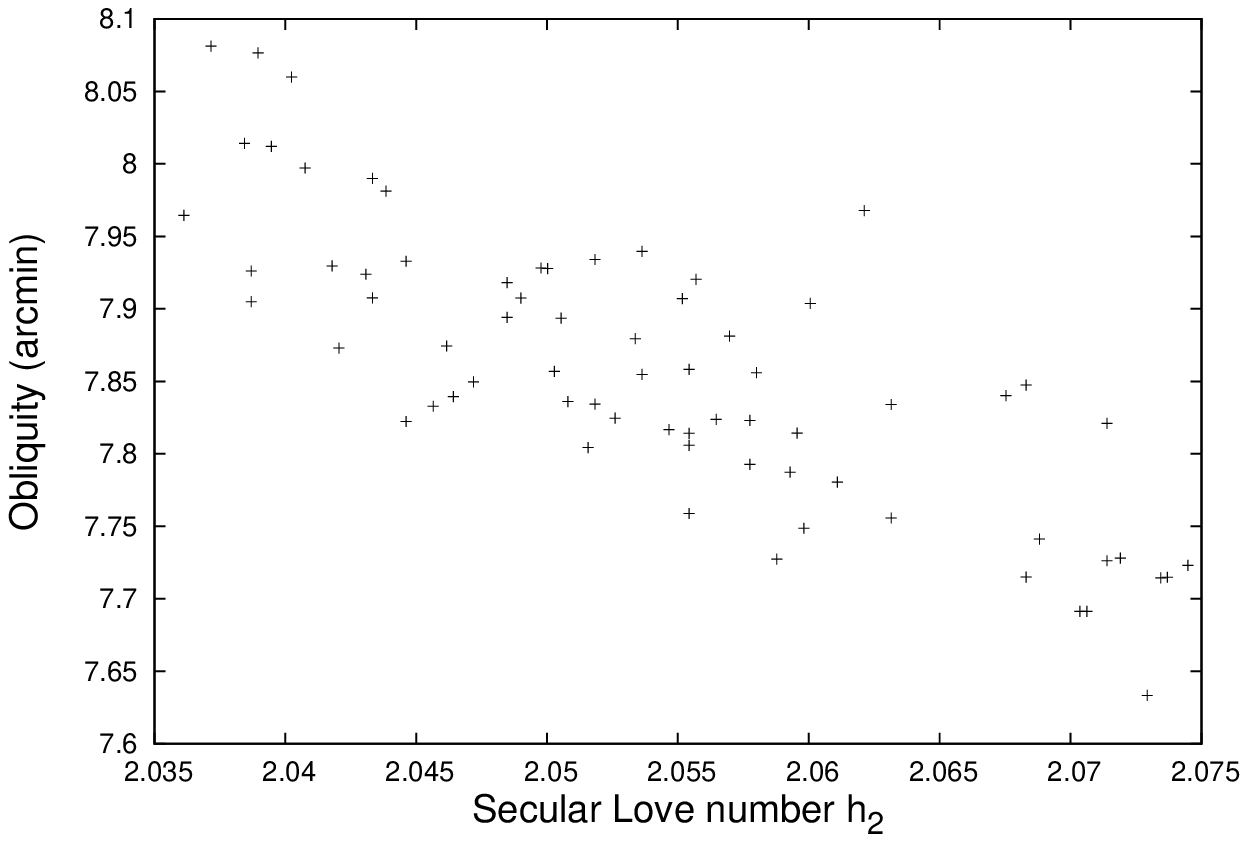} & \includegraphics[width=0.45\textwidth]{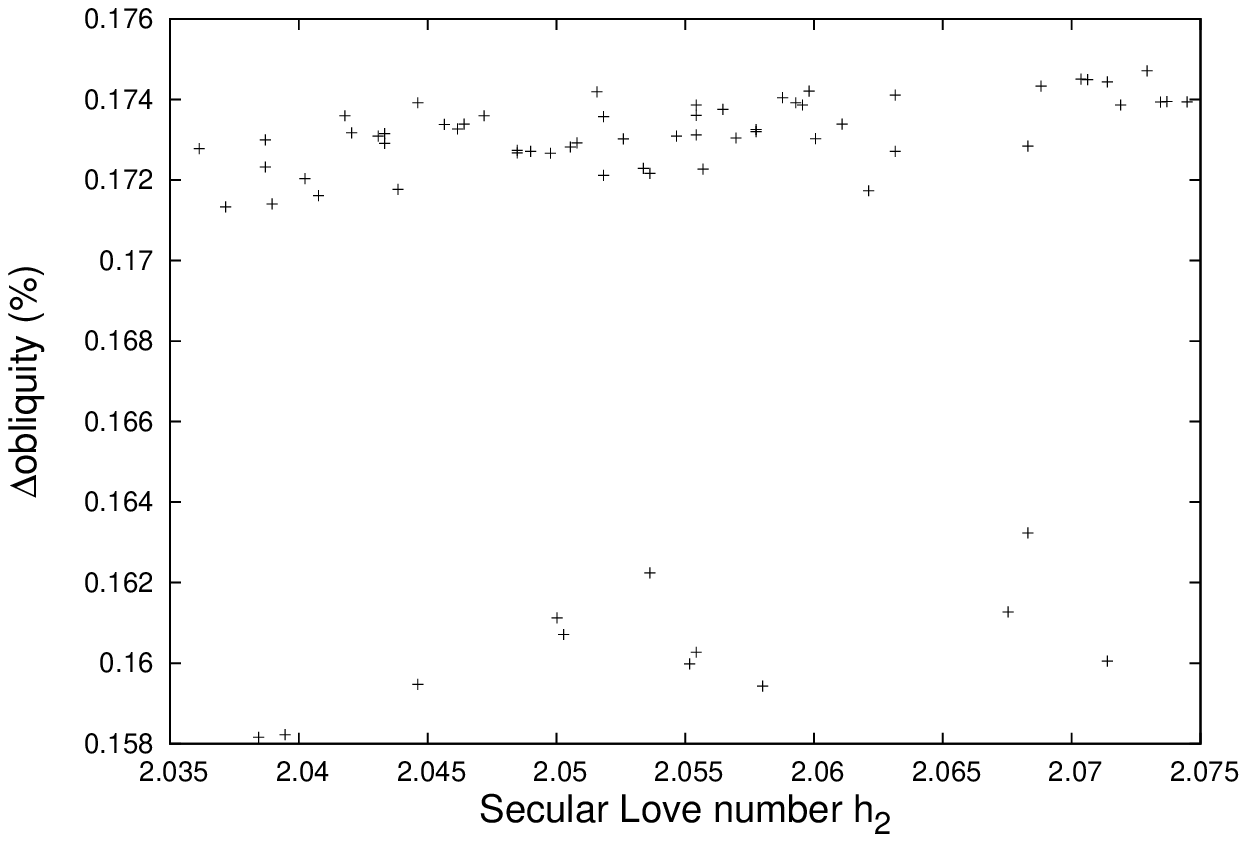}
\end{tabular}
\caption{Amplitude of the main term of the latitudinal librations of the crust of Europa, which is very close to the obliquity (left), and error with the analytical formula (Eq.~\ref{eq:obliqequi}).\label{fig:lati}}
\end{figure}

The Fig.~\ref{fig:longi} represents the first term in longitude, numerically obtained, and the one of the diurnal librations, which comes from an analytical formula (Eq.~\ref{eq:sigma1}),
since it is too small to be numerically detected. The Fig.~\ref{fig:lati} represents the numerically measured obliquity, as a latitudinal libration, with a comparison with the 
analytical formula (Eq.~\ref{eq:obliqequi}). In these formulae, the quantity $(h_2/Q_h)(\chi_1)$ is set to $0$, since I assume the crust to be elastic.

\par We can see that the iterative algorithm gives the obliquity with a difference much less than $1\%$ with respect to the analytical formula. The numerical values of these quantities
should be taken with care, since they are affected by the interior, which is absent here. 

\par
\begin{figure}
\centering
\includegraphics[width=0.45\textwidth]{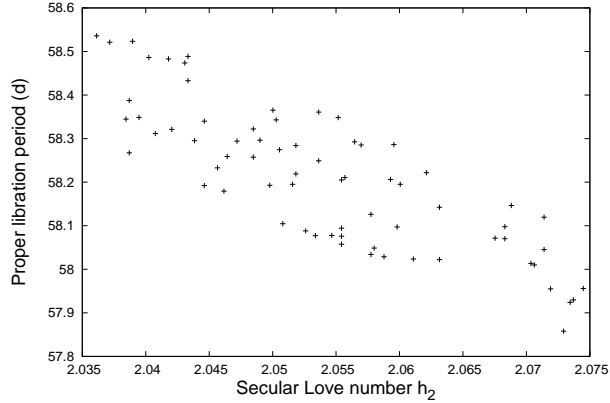}
\caption[Period of the proper longitudinal librations]{Period of the proper longitudinal librations, obtained with the Eq.~\ref{eq:omega02}. This period is to be compared 
with the orbital period, i.e. $3.5512$ d.\label{fig:freqlib}}
\end{figure}

 The Fig.~\ref{fig:freqlib} displays the period of the proper longitudinal librations.

\par Knowing this period allows to check the convergence criterion, i.e. a small $(\omega_0/n)^2$. Since the ratio of the frequencies is the invert of the ratio of the periods, 
in assuming that the proper period is comprised between $57$ and $59$ days, then the ratio $(\omega_0/n)^2$ is comprised between $3.6\times10^{-3}$ and $3.9\times10^{-3}$. This is
another check of the reliability of the convergence, in addition to the comparison between the analytical and numerical amplitudes in the tensor of inertia (Tab.~\ref{tab:I11num}
to \ref{tab:I33num}) and the error on the amplitudes of libration (Fig.~\ref{fig:longi} \& \ref{fig:lati}).

\subsection{A negligible polar motion}

\par
\begin{figure}
\centering
\includegraphics[width=0.45\textwidth]{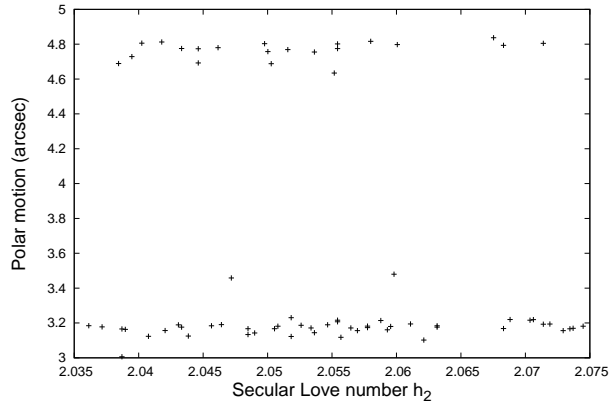}
\caption{Maximal polar motion for the different model-crusts of Europa.\label{fig:wob}}
\end{figure}

 This code considers the 3 degrees-of-freedom, i.e. longitude, obliquity, and polar motion. The polar motion is usually neglected. The Fig.~\ref{fig:wob} 
represents its maximum value over a numerical simulation. It is evaluated from the straightforward formula, which ensues from the modified Andoyer variables:

\begin{equation}
\label{eq:polarmotion}
J = \arccos\left(1-\frac{\xi^2+\eta^2}{2P}\right).
\end{equation}

We can see that the polar motion is at the most $4.9$ arcsec, or $0.082$ arcmin, which is very small with respect to the amplitudes of longitudinal librations, and to the obliquity.
It can be bigger than the variations of the obliquity, but such variations have never been detected for the natural satellites of the giant planets. So, the polar motion remains
negligible with respect to the quantities that can be detected. Neglecting it can then facilitate the computations without a significant loss of accuracy.

\section{Discussion and perspectives}

\par In this study, I have derived the tensor of inertia and the rotational quantities for a viscoelastic and synchronous crust, in assuming small eccentricity and inclination, which is suitable for 
some satellites of the giant planets, harbouring an ocean. I have then proposed and tested an iterative algorithm, devoted to compute consistently the time-varying tensor of inertia and the 
rotational quantities. I have shown in the planar case that the assumed solution is not a fixed point, which casts doubts on its mathematical convergence. Nevertheless, the numerical tests
show that this divergence is acceptable in the elastic case, provided that the proper oscillations are far enough from resonances with forcing terms. The amplitudes are barely affected and 
amplitudes of additional contributions are found. However, the phase shifts appearing in the viscoelastic case diverge too much to be reliable. 

\par This lack of convergence probably illustrates some limitations of the model of hydrostatic shape. Observations prove it to be a reliable first approximation, but many discrepancies
have been found. They could be due to mass anomalies, which could record a past hydrostatic figure, which would correspond to a higher rotation rate \citep[e.g.]{zshlkl2009}.

\par Another issue is that this study does not address the possible super-synchronous rotation of the crust, which could explain features observed at the surface of Europa \citep{gghhmptosgbdcbv1998}.
The tides do not preclude it, under certain conditions \citep{m2015}. Another option to reach a fixed point would be consider a finite amplitude of the proper oscillations at the
equilibrium. \citet{bdg2017} recently showed that this is in some cases required for the equilibrium spin to be stable. Another mystery has been raised by the discovery of a 6$^{\circ}$
phase shift in the diurnal libration of Mimas \citep{trlcrrn2014}, which could be explained only by the presence of a global ocean \citep{c2017,n2017}, while the surface does not present significant
evidence of past tectonic activity.

\par In this study I have worked with the topographic Love number $h_2$, to follow the method by \citep{hla1982,rrc2014}. It is actually possible to use the potential Love number $k_2$ instead,
as I did in \citep{n2017}. A simple analogy between e.g. Eq.~22 in \textit{Ibid.} with $\sigma=0$ and Eq.~\ref{eq:I12freq} shows:

\begin{equation}
\label{eq:h2k2}
k_2\frac{M_{\saturn}R^3}{a^3} = \frac{16}{5}h_2\frac{n^2}{g}\pi\rho X_3.
\end{equation}

\par Using this model of frequency-independent tides in a physically realistic context would need to plug it to a model addressing the rotation of a 3-layer body, with an internal ocean, alike the one recently proposed by \citet{brr2018}.
This would involve additional torques, in particular:

\begin{itemize}

\item The gravitational coupling with the core,

\item The pressure coupling at the core-ocean and ocean-shell boundaries,

\item The coupling with the atmosphere when relevant, i.e. for Titan.

\end{itemize}

\par As shown by previous studies \citep[e.g.]{vbt2013,byv2016}, the first two of these effects not only would add inertia, and may reduce the rotational response of the crust, but they would also add 
proper oscillations, due to the rotation of the crust and the orientation of the velocity field in the fluid. These additional oscillations increase the possibilities of resonant forcings. As
suggested by \citet{rrc2014}, the atmosphere would modulate the annual response, i.e. at the orbital and semi-orbital periods of the parent planet, which are $11.86$ and $5.93$ years for Jupiter, 
and $29.46$ and $14.73$ years for Saturn.

\section*{Acknowledgements}

This study has been funded by the BELgian Science Policy Office (BELSPO). This research used resources of the "Plateforme Technologique de Calcul Intensif (PTCI)"
(http://www.ptci.unamur.be) located at the University of Namur, Belgium, which is supported 
by the F.R.S.-FNRS under the convention No. 2.5020.11. The PTCI is member of the "Consortium des \'Equipements de Calcul Intensif 
(C\'ECI)" (http://www.ceci-hpc.be). I also thank an anonymous reviewer, who raised a critical issue in the calculation of the Love numbers, Michael Efroimsky,
whose comments significantly improved the paper, and Rose-Marie Baland for a careful reading and insightful discussions.

\appendix

\section{Non-Hamiltonian formulation}

\par The rotational dynamics is fully described by the Hamiltonian formalism. The reasons why I also derive non-Hamiltonian equations, that must be equivalent to the previous ones,
is to secure the results, to facilitate comparisons with other studies, and also because they are more convenient to extend the model to more realistic interiors, in particular 
other authors have already expressed the couplings between the crust and the interior \citep[e.g.]{sx1997,bvyk2011,vbt2013,pmhs2016} with this formalism.

\par As explained for instance in \citep{fc1999}, the rotation vector $\vec{\omega}=\omega_1\hat{f}_1+\omega_2\hat{f}_2+\omega_3\hat{f}_3$ represents 3 successive rotations, from the 
inertial frame $(\hat{e}_1,\hat{e}_2,\hat{e}_3)$ to the figure frame $(\hat{f}_1,\hat{f}_2,\hat{f}_3)$:

\begin{enumerate}
	
\item a rotation around the z-axis (here $\hat{e}_3$) of an angle $h$,
	
\item then a rotation around the new, but not final, x-axis of angle $\epsilon$, 
	
\item and finally a rotation around the final z-axis, i.e. $\hat{f}_3$, of an angle $\theta$.
	
\end{enumerate}
Here, the definitions of the Euler angles $h$, $\epsilon$ and $\theta$ are valid only in this subsection. These rotations read mathematically:

\begin{equation}
	\label{eq:kinetic1}
	\left(\begin{array}{c}
	\omega_1 \\
	\omega_2 \\
	\omega_3 
	\end{array}\right) = 
	\left(\begin{array}{c}
	0 \\
	0 \\
	\dot{\theta} 
	\end{array}\right)+R_3(-\theta)\left(\begin{array}{c}
	\dot{\epsilon} \\
	0 \\
	0
	\end{array}\right)+R_3(-\theta)R_1(-\epsilon)\left(\begin{array}{c}
	0 \\
	0 \\
	\dot{h} 
	\end{array}\right),
\end{equation}
i.e.

\begin{eqnarray}
	\omega_1 & = & \dot{\epsilon}\cos\theta+\dot{h}\sin\epsilon\sin\theta, \label{eq:omega1} \\
	\omega_2 & = & -\dot{\epsilon}\sin\theta+\dot{h}\sin\epsilon\cos\theta, \label{eq:omega2} \\
	\omega_3 & = & \dot{\theta}+\dot{h}\cos\epsilon, \label{eq:omega3}
\end{eqnarray}
or

\begin{eqnarray}
	\dot{h} & = & \frac{\omega_1\sin\theta+\omega_2\cos\theta}{\sin\epsilon}, \label{eq:doth} \\
	\dot{\epsilon} & = & \omega_1\cos\theta-\omega_2\sin\theta, \label{eq:dotepsilon} \\
	\dot{\theta} & = & \omega_3-\frac{\omega_1\sin\theta+\omega_2\cos\theta}{\tan\epsilon}. \label{eq:dottheta}
\end{eqnarray}

\par The formulae are the same as the ones present in \citep{bnt1995,htn2011}, but are given with a different sign in numerous other studies, \citep[e.g.]{wbyrd2001},
probably because of a different sign convention. The Eq.(\ref{eq:doth}) and (\ref{eq:dottheta}) show that the set of Euler angles $(h,\epsilon,\theta)$ present a 
virtual singularity when the kind-of-obliquity $\epsilon$ is small. To bypass this problem I define these other variables:

\begin{eqnarray}
	\xi  & = & 2\sin\frac{\epsilon}{2}\sin h, \label{eq:xi} \\
	\eta & = & 2\sin\frac{\epsilon}{2}\cos h, \label{eq:eta} \\
	p    & = & h+\theta,	\label{eq:p}
\end{eqnarray}
yielding

\begin{eqnarray}
	\dot{\xi}  & = & \omega_1\left(\cos\theta\cos\frac{\epsilon}{2}\sin h+\frac{\sin\theta\cos h}{\cos\left(\epsilon/2\right)}\right)
                        +\omega_2\left(\frac{\cos\theta\cos h}{\cos\left(\epsilon/2\right)}-\sin\theta\cos\frac{\epsilon}{2}\sin h\right), \label{eq:dotxi} \\
	\dot{\eta} & = & \omega_1\left(\cos\theta\cos\frac{\epsilon}{2}\cos h-\frac{\sin\theta\sin h}{\cos\left(\epsilon/2\right)}\right)
                        -\omega_2\left(\sin\theta\cos\frac{\epsilon}{2}\cos h+\frac{\cos\theta\sin h}{\cos\left(\epsilon/2\right)}\right), \label{eq:doteta} \\
	\dot{p}    & = & \omega_3+\left(\omega_1\sin\theta+\omega_2\cos\theta\right)\tan\frac{\epsilon}{2}. \label{eq:dotp}
\end{eqnarray}

\par The Eq.(\ref{eq:dotxi}) to (\ref{eq:dotp}) are the 3 kinematic equations ruling the behavior of the Euler angles, I now need to write down the 3 dynamical equations
ruling the components of the rotation vector $(\omega_1,\omega_2,\omega_3)$.

\par We have classically

\begin{equation}
\label{eq:dynG}
\frac{d\vec{G}}{dt} = \vec{\Gamma}-\vec{\omega}\times\vec{G},
\end{equation}
$\vec{\Gamma}=\Gamma_1\hat{f}_1+\Gamma_2\hat{f}_2+\Gamma_3\hat{f}_3$ being the total torque acting on the shell, and the term $-\vec{\omega}\times\vec{G}$ compensates 
the fact that I express $\vec{G}$ in the figure frame $(\hat{f}_1,\hat{f}_2,\hat{f}_3)$, that is not inertial. In this study, $\vec{\Gamma}$ is the gravitational pull of the parent planet.

\par A derivation of $\vec{\Gamma}$ is given in \citep[Chap.~5]{md1999}, itself inspired from \citep{m1936,r1937,r1940}. The perturbing potential from which $\Gamma$ derives reads

\begin{equation}
\label{eq:potentielpert}
V = -\frac{\mathcal{G}M_pm}{r}-\frac{\mathcal{G}M_p}{2r^3}\left(I_{11}+I_{22}+I_{33}-3I\right),
\end{equation}
where $I$ is the moment of inertia about the line OP, O and P being defined as before, i.e. O is the barycenter of the shell, and P the barycenter of the perturber.
From

\begin{equation}
\label{eq:mominteriaOP}
I = I_{11}x^2+I_{22}y^2+I_{33}z^2+2I_{12}xy+2I_{13}xz+2I_{23}yz
\end{equation}
I get, in using the relation $x^2+y^2+z^2=1$:

\begin{equation}
\label{eq:potentielpert2}
V = -\frac{\mathcal{G}M_pm}{r}-\frac{\mathcal{G}M_p}{2r^3}f
\end{equation}
with

\begin{equation}
\label{eq:thef}
f = (I_{22}+I_{33}-2I_{11})x^2+(I_{11}+I_{33}-2I_{22})y^2+(I_{11}+I_{22}-2I_{33})z^2-6I_{12}xy-6I_{13}xz-6I_{23}yz.
\end{equation}

From 

\begin{eqnarray}
F_1 & = & -\frac{\partial V}{\partial x}, \label{eq:f1} \\
F_2 & = & -\frac{\partial V}{\partial y}, \label{eq:f2} \\
F_3 & = & -\frac{\partial V}{\partial z}, \label{eq:f3} \\
\Gamma_1 & = & zF_2-yF_3, \label{eq:g1} \\
\Gamma_2 & = & xF_3-zF_1, \label{eq:g2} \\
\Gamma_3 & = & yF_1-xF_2, \label{eq:g3}
\end{eqnarray}

I get

\begin{eqnarray}
\Gamma_1 & = & 3\frac{\mathcal{G}M_p}{r^3}\left((I_{33}-I_{22})yz+I_{23}(y^2-z^2)+I_{13}xy-I_{12}xz\right), \label{eq:Gamma1} \\
\Gamma_2 & = & 3\frac{\mathcal{G}M_p}{r^3}\left((I_{11}-I_{33})xz+I_{13}(z^2-x^2)+I_{12}yz-I_{23}xy\right), \label{eq:Gamma2} \\
\Gamma_3 & = & 3\frac{\mathcal{G}M_p}{r^3}\left((I_{22}-I_{11})xy+I_{12}(x^2-y^2)+I_{23}xz-I_{13}yz\right). \label{eq:Gamma3}
\end{eqnarray}

\par The coordinates of the unit vector $(x,y,z)$ pointing to the perturber in the body frame are obtained from its coordinates in the
inertial frame $(x_i,y_i,z_i)$ after 3 rotations:

\begin{equation}
\label{eq:3rotations}
\left(\begin{array}{c}
x \\
y \\
z \end{array}\right)
= R_3(-\theta)R_1(-\epsilon)R_3(-h)\left(\begin{array}{c}
x_i \\
y_i \\
z_i \end{array}\right),
\end{equation}
while the rotation vector $\vec{\omega}$ comes from the angular momentum (Eq.~\ref{eq:defomega}). Finally, our 6 variables are 
$(\xi,\eta,p,G_1,G_2,G_3)$, and their evolution is known in integrating their time derivatives defined by the equations
(\ref{eq:dotxi}) to (\ref{eq:dotp}) and (\ref{eq:Gamma1}) to (\ref{eq:Gamma3}).

\bsp	
\label{lastpage}
\end{document}